\documentstyle[epsf,12pt]{article}

\def\bc{\begin{center}}
\def\ec{\end{center}}
\def\be{\begin{equation}}
\def\ee{\end{equation}}
\def\myappendix{\par
 \setcounter{section}{0}
 \setcounter{subsection}{0}

 \setcounter{equation}{0}
 \setcounter{table}{0}
 \def\appendixname{Appendix}
 \def\appesection{\setcounter{equation}{0}\section}
 \def\@thesection{\Alph{section}}
 \def\thesection{\appendixname\hskip 1.10ex\Alph{section}}
 \def\thesubsection{\@thesection.\arabic{subsection}}
 \def\theequation{\@thesection.\arabic{equation}}
 \def\thetable{\@thesection.\arabic{table}}}

\newcommand{\beq}{\begin{equation}}
\newcommand{\eeq}{\end{equation}}
\newcommand{\beqn}{\begin{eqnarray}}
\newcommand{\eeqn}{\end{eqnarray}}
\newcommand{\beqns}{\begin{eqnarray*}}
\newcommand{\eeqns}{\end{eqnarray*}}
\newcommand{\mfrac}[2]{\frac{\textstyle #1}{\textstyle #2}}

\def\vdir{v\kern-7.8pt\Big{/}}
\def\pdir{p\kern-7.8pt\Big{/}}
\def\ddir{D\kern-7.8pt\Big{/}}

\begin{document}
\pagestyle{empty} 
\vspace{-0.6in}
\begin{flushright}
FTUV 98/2 - IFIC 98/2
\end{flushright}
\vskip 0.2 cm
\centerline{\LARGE{Calculation of the continuum--lattice HQET matching}}
\centerline{\LARGE{for
the complete basis of four--fermion operators:}}
\centerline{\LARGE{reanalysis of the $B^{0}$-$\bar{B}^{0}$ mixing.}}
\vskip 1.4cm
\centerline{\bf{V. Gim\'enez and
J. Reyes\footnote{Gobierno Vasco predoctoral Fellow}}}
\vskip 0.5cm
\centerline{Dep. de Fisica Teorica and IFIC, Univ. de Valencia,}
\centerline{Dr. Moliner 50, E-46100, Burjassot, Valencia, Spain.} 
\abstract{
In this work, we find the expressions of continuum HQET 
four-fermion operators in terms of lattice operators in perturbation theory.
To do so, we calculate the one--loop continuum--lattice HQET matching for the 
complete basis of $\Delta B=2$ and $\Delta B=0$ operators 
(excluding penguin diagrams), extending and completing previous studies. 
We have also corrected some errors in previous evaluations of the matching for
the operator $O_{LL}$. 
Our results are relevant to the 
lattice computation of the values of unknown hadronic matrix elements which 
enter in many very important theoretical predictions in $B$--meson 
phenomenology: $B^{0}$-$\bar{B}^{0}$ mixing, $\tau_{B}$ and $\tau_{B_{s}}$ 
lifetimes, SUSY effects in $\Delta B=2$ transitions and the 
$B_{s}$ width difference $\Delta \Gamma_{B_{s}}$.
We have reanalyzed our lattice data for the $B_{B}$ parameter of the 
$B^{0}$-$\bar{B}^{0}$ mixing on 600 lattices of size $24^{3}\times 40$
at $\beta=6.0$ computed with the SW-Clover and HQET lattice actions. 
We have used the correct lattice--continuum matching
factors and boosted perturbation theory with tadpole improved heavy--light operators to
reduce the systematic error in the evaluation of the renormalization constants.
Our best estimate of the renormalization scale independent $B$--parameter is
$\hat{B}_{B} = 1.29\, \pm\, 0.08\, \pm\, 0.06$, where the first error
is statistical and the second is systematic coming from the uncertainty 
in the determination of the renormalization constants. Our result 
is in good agreement with previous results obtained by extrapolating Wilson data.
As a byproduct, we also obtain the complete one--loop anomalous dimension 
matrix for four--fermion operators in the HQET.
}
\vfill
\centerline{PACS: 11.15.Ha, 12.38.Gc, 12.38.Bx, 12.39.Hg, 13.25.Hw, 14.40.Nd}
\vfill
\eject
\pagestyle{empty}\clearpage
\setcounter{page}{1}
\pagestyle{plain}
\newpage 
\pagestyle{plain} \setcounter{page}{1}
\section{Introduction and Motivation.}
\label{motiv}
$B$--hadron decays are a very important source of information on the physics
of the standard model (SM) and beyond. In many important cases, long
distance strong contributions to these processes can be separated into matrix 
elements of local operators. Lattice QCD can then be used to compute these 
non--perturbative parameters from first principles. 
A list of some four--fermion operators relevant to $B$--meson phenomenology
is the following:
%\begin{itemize}
%\item the operator $O_{LL}=\bar b\, \gamma^{\mu}\, (1-\gamma_{5})\, q\, 
%              \bar b\, \gamma_{\mu}\, (1-\gamma_{5})\, q$ ( $q$ denotes a light
%              quark $u$, $d$ or $s$), which determines the
%              theoretical prediction of the $B^0$-$\bar{B}^{0}$ mixing
%              in the SM, together with 
%$O^{S}_{LL(RR)}=\bar b\, (1\mp\gamma_{5})\, q\, \bar b\, (1\mp\gamma_{5})\, q$,
%$O_{RR}$ and $O^{S}_{LR}$, with obvious notation, parameterize SUSY effects 
%in $\Delta B=2$ transitions \cite{susy}. Furthermore, $O_{LL}$ and $O^{S}_{LL}$ determine
%the $B_{s}$ width difference $\Delta \Gamma_{B_{s}}$ \cite{deltabs};
%\item $Q_{LL}=\bar b\, \gamma^{\mu}\, (1-\gamma_{5})\, q\, 
%              \bar q\, \gamma_{\mu}\, (1-\gamma_{5})\, b$, together with 
%              $Q^{S}_{LR}$, parameterize
%              spectator effects in the $\tau_{B}$ and $\tau_{B_{s}}$ lifetimes
%              \cite{taubbs};
%\item finally, the corresponding operators with a $t^{a}$ (the generator of the
%SU(3) group) insertion: 
%$Ot_{LL}$, $Ot^{S}_{LL}$ and $Ot^{S}_{LR}$ contribute to SUSY effects in 
%$\Delta B=2$ transitions \cite{susy}, and $Qt_{LL}$ and $Qt^{S}_{LR}$ to spectator effects 
%in the B-meson lifetimes \cite{taubbs}.                            
%\end{itemize}
\begin{eqnarray}
\label{allopero}
\mbox{\rm \bf $\Delta B=2$ operators}&&\nonumber\\
O_{LL(RR)} &=& \bar b\, \gamma^{\mu}\, (1\mp\gamma_{5})\, q\, 
 \bar b\, \gamma_{\mu}\, (1\mp\gamma_{5})\, q\nonumber\\
O^{S}_{LL(RR)} &=& \bar b\, (1\mp\gamma_{5})\, q\, 
 \bar b\,  (1\mp\gamma_{5})\, q\nonumber\\
O_{LR(RL)} &=& \bar b\, \gamma^{\mu}\, (1\mp\gamma_{5})\, q\, 
 \bar b\, \gamma_{\mu}\, (1\pm\gamma_{5})\, q\nonumber\\
O^{S}_{LR(RL)} &=& \bar b\, (1\mp\gamma_{5})\, q\, 
 \bar b\,  (1\pm\gamma_{5})\, q\\
\label{alloperq}
\mbox{\rm \bf $\Delta B=0$ operators}&&\nonumber\\ 
Q_{LL(RR)} &=& \bar b\, \gamma^{\mu}\, (1\mp\gamma_{5})\, q\, 
 \bar q\, \gamma_{\mu}\, (1\mp\gamma_{5})\, b\nonumber\\
Q^{S}_{LL(RR)} &=& \bar b\, (1\mp\gamma_{5})\, q\, 
 \bar q\,  (1\mp\gamma_{5})\, b\nonumber\\
Q_{LR(RL)} &=& \bar b\, \gamma^{\mu}\, (1\mp\gamma_{5})\, q\, 
 \bar q\, \gamma_{\mu}\, (1\pm\gamma_{5})\, b\nonumber\\
Q^{S}_{LR(RL)} &=& \bar b\, (1\mp\gamma_{5})\, q\, 
 \bar q\,  (1\pm\gamma_{5})\, b
\end{eqnarray}
where $q$ denotes a light quark $u$, $d$ or $s$.  
The corresponding operators with a $t^{a}$ (the generator of the
SU(3) group) insertion, denoted by $Ot_{LL}$ \ldots etc, will also be considered.
 
As is well known, $O_{LL}$ determines the  theoretical prediction of the 
$B^0$-$\bar{B}^{0}$ mixing in the SM. Moreover, $O_{LL}$ together with 
$O^{S}_{LL(RR)}$,
$O_{RR}$ and $O^{S}_{LR}$, parameterize SUSY effects 
in $\Delta B=2$ transitions \cite{susy}. $O_{LL}$ and $O^{S}_{LL}$ 
determine the $B_{s}$ width difference $\Delta \Gamma_{B_{s}}$ \cite{deltabs}.
The $\Delta B=0$ operators $Q_{LL}$ and $Q^{S}_{LR}$, parameterize
spectator effects in the $\tau_{B}$ and $\tau_{B_{s}}$ lifetimes \cite{taubbs}.
For the four--fermion operators with a $t^{a}$ insertion: 
$Ot^{S}_{LL}(RR)$ and $Ot^{S}_{LR}$ contribute to SUSY effects in 
$\Delta B=2$ transitions \cite{susy}, and $Qt_{LL}$ and $Qt^{S}_{LR}$ to spectator effects 
in the B-meson lifetimes \cite{taubbs}.                            

Our aim is to non-perturbatively evaluate the matrix elements 
between $B$--meson states of all the operators in eqs.(\ref{allopero}) and (\ref{alloperq}),
using lattice simulations of the $b$--quark in the HQET. These non perturbative
parameters can also be measured in principle with propagating $b$ quarks by 
simulating in the charm mass region and extrapolating to the $b$ quark mass.  
The procedure to perform the transition from QCD to lattice HQET, 
a combination of analytic and numerical calculations, can be split into
the following four steps:
\begin{description}
\item[{\bf First Step:}{\it\,\,  the continuum QCD -- HQET matching.}]\mbox{ }\\
The QCD operators are expressed as linear combinations of
HQET ones in the continuum at a given high scale, say, $\mu=m_{b}$.
\item[{\bf Second Step:}{\it\,\,  running down to $\mu=a^{-1}$ in the HQET.}]\mbox{ }\\
The HQET operators obtained in step 1 at the high scale $\mu=m_{b}$ are 
evolved down to a lower scale $\mu=a^{-1}$, appropriate for lattice simulations,
using the HQET NLO renormalization group equations. 
\item[{\bf Third Step:}{\it\,\, continuum--lattice HQET matching.}]\mbox{ }\\
Having obtained the continuum HQET operators at the scale $\mu=a^{-1}$,
they are expressed as a linear combination of lattice 
HQET operators at this scale. The procedure is very similar to step 1: 
two amplitudes are matched at one-loop order, one 
in the continuum HQET and the other in the lattice HQET. 
\item[{\bf Fourth Step:}{\it\,\, lattice computation of the matrix elements.}]\mbox{ }\\
The matrix elements of the lattice HQET operators in Step 3, are measured by 
Monte Carlo numerical simulations on the lattice. 
Using these values and the chain of matching 
equations in steps 1 to 3, the matrix elements of the continuum QCD operators 
relevant to phenomenology can be determined.
\end{description}

In this work, we deal with step 3, i.e.~we find the expressions of
the continuum HQET operators in terms of lattice HQET ones using continuum 
and lattice perturbation theory at one loop. 
The calculation of steps 1 and 2, and the 
numerical simulation in step 4 are in progress and will be published elsewhere.

We would like to stress that a non 
perturbative determination of the lattice renormalization constants would be
preferable because at the values of the coupling at which the simulations are
performed, the one--loop perturbative corrections are not small, due 
to the appearance of tadpole diagrams. In principle, the Martinelli's {\it et
al} \cite{npr} method for non perturbative renormalization may be
utilized to avoid the problems with perturbation theory. Unfortunately,
however, the application of the former to the HQET is not straightforward \cite{za}.
%due to the presence of renormalon singularities in this theory \cite{za}.  
The key observation here is that HQET lattice simulations suggest that 
the finite heavy quark propagator in the Landau gauge has the form \cite{labar}
\be
S(\vec{x},t)\, =\, \delta(\vec{x})\, \theta(t)\, A(t)\, e^{-(\lambda - \delta
\bar{m})\, t}
\ee
where $A(t)$ is a smooth unknown function of $t$, $\lambda$ is a constant and 
$\delta \bar{m}$ is the
residual mass (also called the HQET mass counter--term). The latter is
needed to remove the linear divergence in $\lambda$.
% corresponding to an ultraviolet renormalon in dimensional regularization.
 On the one hand, the
difference $m_{S}\equiv \lambda - \delta \bar{m}$ should be small in order to
reduce $O(m_{S} a)$ terms. On the other hand, however, a too small $m_{S}$
implies that the heavy quark propagator goes to a constant. This fact makes its
numerical Fourier transform, a necessary ingredient in the  
Martinelli's {\it et al} method, very problematic due to the appearance of huge
finite time effects of $O(e^{-m_{S}\, T})$, where $T$ is the lattice time 
length. Therefore, the non perturbative calculation of renormalization
constants in the HQET is difficult and requires more careful studies.

The paper is organized as follows. We begin by introducing the lagrangian
of the HQET in the continuum and its discretized version which we will use in
lattice simulations. The corresponding Feynman rules are derived for both heavy quarks
and antiquarks. In section 3, we give the lattice lagrangian for light quarks
and the corresponding Feynman rules. The strategy of the calculation 
is presented in section 4. In this section, our renormalization prescription is
discussed in detail. Moreover, we study the subtle point of the 
presence of evanescent operators and their effect on the matching and
higher--order calculations. 
Then, we present the calculation details of
the Feynman diagrams in the continuum and on the lattice HQET using several renormalization
schemes. In section 6, our results for the one--loop continuum--lattice matching 
are given and compared with previous partial calculations. 
We also discuss the effects of a change of the continuum renormalization scheme. 
Then we reanalyze the $B^{0}$-$\bar{B}^{0}$ mixing using the correct matching
factors. Finally, we present our conclusions in section 8.
In addition, we include one appendix where the relevant formula used in our
calculation are collected.

\section{Feynman rules for the continuum and discre\-ti\-zed HQET
la\-gran\-gians.}

The Minkowski continuum lagrangian for a field $h(x)$ which annihilates
a static heavy quark and for a field $\tilde{h}(x)$ which annihilates
a static heavy antiquark is
%\be
%{\cal L\}_{HQET}\,=\, \bar{h}(x)\, i\, D_{0}\, h(x)\, +\,
%\bar{\tilde{h}}(x)\, i\, D_{0}\, \tilde{h}(x)
%\ee
\be 
\label{laghqetc}
{\cal L}_{HQET}^{cont}(x)\,=\, h^{\dagger}(x)\, i\, D_{0}\, h(x)\, +\,
\tilde{h}(x)\, i\, D_{0}\, \tilde{h}^{\dagger}(x)
\ee
where $i\, D_{\mu}$ is the covariant derivative
$i\, \partial_{\mu}\, +\, g\, t^{a}\, A_{\mu}^{a}(x)$ with $A^{a}_{\mu}(x)$
the gluon field and $t^{a}=\lambda^{a}/2$ the generators of the SU(3) color
group normalized by $tr(t^{a} t^{b})=\delta^{a b}/2$.
A very important property of the HQET fields $h$ and $\tilde{h}$
is the fact that $\gamma_{0}\, h\, =\, h$ and $\tilde{h}\, \gamma_{0}\,  =\,
-\, \tilde{h}$. We will make
extensive use of these properties in our calculation.

On the lattice, in Euclidean space, we have first to choose a discretization 
for the continuum
covariant derivative $D_{\mu}$, which has the correct continuum limit.
Obviously there are many equivalent possibilities.
The point is that the discretization we use in the
perturbative calculation of the lattice operators must be the same as the one
with which we perform the numerical simulation. Otherwise, the mismatch will generate
spurious uncontrolled finite contributions to the final results.
Our convention is to discretize backward
in time, as first proposed by Eichten \cite{eh1}, so that the
lattice HQET action is
\begin{eqnarray} \label{laghqetl}
S^{HQET} &=& a^{3}\, \sum_{n}\, \left\{\, h^{\dagger}(n)\,
\left[\, h(n)\, -\, U^{\dagger}_{4}(n-\hat{4})\, h(n-\hat{4})\, \right]\right.
\nonumber\\
&+& \left. \tilde{h}(n)\,
\left[\, U_{4}(n)\, \tilde{h}^{\dagger}(n+\hat{4})\, -\, 
\tilde{h}^{\dagger}(n)\right] \right\}
\end{eqnarray}
where $n$ denotes a lattice site, $\hat{4}$ is the unit vector in
the Euclidean time direction and $U$ is defined in eq.(\ref{defu}) bellow. 

\begin{table}[t]

\begin{center}
\begin{tabular}{||l|r|r||}
\hline
\hline
\mbox{\rm Quark}&\multicolumn{1}{|c|}{\mbox{\rm Minkowski HQET}}
&\multicolumn{1}{|c||}{\mbox{\rm Lattice HQET}}\\
\hline
\mbox{}&\mbox{}&\mbox{}\\[-9.5pt]
\mbox{\rm Propagator}&$\mfrac{i}{p^{0}\,+\, i\, \epsilon}$&$
\mfrac{a}{1\, +\, \epsilon\, -\, e^{-i\, p_{4}\, a}}$\\
\mbox{}&\mbox{}&\mbox{}\\[-9.5pt]
\mbox{\rm quark-gluon vertex}&$+\, i\, g\, t^{a}_{\beta\, \alpha}\, g^{\mu 0}$&$
-\, i\, g\, t^{a}_{\beta\, \alpha}\, \delta_{\mu 4}\,
e^{-\, i\, p_{4}'\, a}$\\
\mbox{}&\mbox{}&\mbox{}\\[-9.5pt]
\mbox{\rm seagull vertex}&{\rm does not exist}&
$-\, 1/2\, a\, g^{2}\, \delta_{\mu 4}\, \delta_{\nu 4}\,
\{\, t^{a} , t^{b}\, \}_{\beta\, \alpha}\,
e^{-\, i\, p_{4}'\, a}$\\
\mbox{}&\mbox{}&\mbox{}\\[-9.5pt]
\hline
\hline
\mbox{\rm Antiquark}&\multicolumn{1}{|c|}{\mbox{\rm Minkowski HQET}}
&\multicolumn{1}{|c||}{\mbox{\rm Lattice HQET}}\\
\hline
\mbox{}&\mbox{}&\mbox{}\\[-9.5pt]
\mbox{\rm Propagator}&$\mfrac{i}{p^{0}\,+\, i\, \epsilon}$&$
\mfrac{a}{1\, +\, \epsilon\, -\, e^{-i\, p_{4}\, a}}$\\
\mbox{}&\mbox{}&\mbox{}\\[-9.5pt]
\mbox{\rm quark-gluon vertex}&$-\, i\, g\, t^{a}_{\alpha\, \beta}\, g^{\mu 0}$&$
+\, i\, g\, t^{a}_{\alpha\, \beta}\, \delta_{\mu 4}\,
e^{-\, i\, p_{4}'\, a}$\\
\mbox{}&\mbox{}&\mbox{}\\[-9.5pt]
\mbox{\rm seagull vertex}&{\rm does not exist}&
$+\, 1/2\, a\, g^{2}\, \delta_{\mu 4}\, \delta_{\nu 4}\,
\{\, t^{a} , t^{b}\, \}_{\alpha\, \beta}\,
e^{-\, i\, p_{4}'\, a}$\\[3pt]
\hline
\hline
\end{tabular}
\end{center}
 \caption{Feynman rules for the continuum Minkowski HQET and the discretized
Euclidean HQET. Our conventions are discussed in the text.}
\label{tabfeynruleshqet}
\end{table}

From eqs.(\ref{laghqetc}) and (\ref{laghqetl}), it is easy to
find the Feynman rules which are collected in Table \ref{tabfeynruleshqet}.
Some remarks are in order here. To perform a weak--coupling expansion in terms
of $g$ of the lattice lagrangian, we have parameterized the link variables as
\be \label{defu}
U^{\dag}_{\mu}(x) = e^{-i a g\, A^{a}_{\mu}(x)\, t^{a}}
\ee
where $A^{a}_{\mu}(x)$ is the gluon field and $a$ is the lattice spacing. 
In Table \ref{tabfeynruleshqet}, the propagator 
corresponds to a quark (antiquark) of momentum $p$, and a color conservation
is understood.  The quark--gluon vertex
corresponds to an incoming quark (antiquark) of color $\alpha$ and momentum $p$,
outgoing quark (antiquark) of color $\beta$ and momentum $p'$ and incoming gluon 
of color index $a$ and Lorentz index $\mu$. The lattice seagull vertex, which
will give rise to tadpole diagrams, is defined as the quark--gluon vertex except for 
an additional incoming gluon with color index $b$ and Lorentz index $\nu$.

\section{The light--quark Wilson and Clo\-ver ac\-tions.}

The operators in eqs.(\ref{allopero}) and (\ref{alloperq}) contain both a heavy and a light quark.
The dynamics of the former is governed by the HQET lagrangian described in 
section 2. In this section, we consider the dynamics of the light degrees of
freedom.

In the continuum, we use the standard QCD lagrangian for light quarks, which
Feynman rules can be found in many text books (see, for example, ref.\cite{tarrach}). 

On the lattice, and in our numerical simulations, light quark propagation is described 
through the Wilson action \cite{wilson}  
\begin{eqnarray} \label{wilson}
S^{W} & =& S_{G}[U]\, +\, a^{4}\, \sum_{n,\mu}\, \left\{- \mfrac{1}{2\, a}\,
\left[\, \bar{\psi}(n)\, ( r\, -\, \gamma_{\mu} )\, U_{\mu}(n)\,
\psi(n+\hat{\mu})\right. \right. \nonumber\\
&+& \left. \left. \bar{\psi}(n+\hat{\mu})\, ( r\, +\, \gamma_{\mu} )\,
U^{\dagger}_{\mu}(n)\, \psi(n)\, \right]\, \right\}
\, +\,  a^{4}\, \sum_{n}\, \left[\, \bar{\psi}(n)\,
( m_{0}\, +\, \mfrac{4 r}{a} )\, \psi(n)\right]
\end{eqnarray}
where $S_{G}[U]$ is the standard Wilson plaquette action \cite{wilson},
$r$ is the Wilson parameter and $m_{0}$ is the bare light fermion mass.

An important source of systematic error in lattice simulations is the
finiteness of the lattice spacing, $a$. It is well--known that hadronic matrix elements 
computed with the Wilson action differ from the corresponding continuum ones by terms of 
$O(a)$. Some methods have been devised to reduce
the cut--off dependence of the lattice hadronic matrix elements involving
Wilson fermions. In this paper, we will apply the method of improvement
first proposed by B.~Sheikholeslami and R.~Wohlert (SW) long ago \cite{sw}, which
has been deeply studied in ref.\cite{swclover}. The main idea of this technique
is to use the freedom we have in the definition of the discretized quark
action to eliminate all terms of $O(a)$ from the lattice matrix
elements, by adding some irrelevant $O(a)$ terms to the action 
(\ref{wilson}) and redefining the lattice fields $\psi$.

The first recipe is to perform all calculations with
the so--called SW-Clover action, given by
\be \label{clover}
S^{I}\, =\, S^{W}\, -\, a^{4}\, \sum_{n,\mu,\nu}\,
\left[\, i\, g\, \mfrac{a r}{4}\, \bar{\psi}(n)\, \sigma_{\mu \nu}
\, P_{\mu \nu}(n)\, \psi(n)\, \right]
\ee
where $P_{\mu \nu}(n)$ is the discretized field strength tensor defined by
\be
P_{\mu \nu}\, =\, \mfrac{1}{4 a^{2}}\, \sum_{i=1}^{4}\,
\mfrac{1}{2 i g}\, (U_{i} - U_{i}^{\dagger})
\ee
and the sum is over the four plaquettes in the $\mu$--$\nu$ plane, stemming
from the point $n$ and taken counterclockwise.

The second recipe is the rule for constructing improved operators:
the quark fields in lattice operators must be rotated
\be \label{psirot}
%\psi' \, =\, \left( 1\, -\, a\, \mfrac{r}{2}\, \vec{\ddir}\,
%\right)\, \psi
\psi' \, =\, \left( 1\, -\, a\, \mfrac{r}{2}\, \left[\, z\, \vec{\ddir}\,
-\, (1-z)\, m_{0}\, \right]\right)\, \psi
\ee
where $\vec{\ddir}$ is the symmetric lattice covariant derivative and $z$
is an arbitrary real number.

It can be demonstrated that on--shell lattice matrix 
elements of improved ope\-ra\-tors com\-pu\-ted with the SW--Clo\-ver ac\-tion 
do not contain $O((g^{2})^{n}\, a\, \log^{n} a)$ terms for any value of $z$.
We will take $z=1$
so that the rotated operator do not depend on $m_{0}$. The reason for this
choice is that in this case perturbative calculations are more accurate 
\cite{swclover}.
 
The same method can be applied to the Eichten's action for heavy quarks in
eq.(\ref{laghqetl}). In ref.\cite{bp}, it is shown, however, that for $O(a)$ 
improvement of on--shell matrix elements, no
modification of the static quark propagator is needed. Therefore,
in order to eliminate $O(a)$ terms from on--shell matrix elements of
heavy--light operators, we have to perform the calculation with the 
SW-Clover action and rotate the light quark fields only.
For heavy quarks, the usual unimproved Eichten's action (\ref{laghqetl})
can be used.

For practical reasons, instead of rotating the fields in the numerical
simulation, a rotation of the light--quark propagator is done \cite{frezzoti}.
This technique is equivalent to a new definition
of the lattice quark fields and automatically improves the operators.
Since this method is much easier to be implemented on the computer
than the improvement with rotated fields, we will apply it in this paper.

\begin{table}[t]
\begin{tabular}{||c|r||}
\hline
\hline
\mbox{}&\mbox{}\\[-10pt]
\mbox{\rm Propagator}&$\left[ 1/a\, \sum_{\mu}\,
\left( i\, \gamma_{\mu}\, \sin(p_{\mu} a) \right)\, +\,
\left( m_{0}\, +\, 2\, \mfrac{r}{a}\, \sum_{\mu}\, \sin^{2}((p_{\mu} a)/2)
\right)\, \right]^{-1}$\\[7pt]
\hline
\mbox{}&\mbox{}\\[-10pt]
\mbox{\rm Wilson quark-gluon vert.}&
$(-i\, g)\, \mfrac{\lambda^{a}_{\beta \alpha }}{2}\,
\left[\, \gamma_{\rho}\, \cos ( (p+p')_{\rho} \mfrac{a}{2} )\, -\,
i\, r\, \sin ( (p+p')_{\rho} \mfrac{a}{2} )\, \right]\, e^{-i p'_{\rho} 
\frac{a}{2}}\, e^{i p_{\rho}\frac{a}{2}}$\\[7pt]
\hline
\mbox{}&\mbox{}\\[-10pt]
\mbox{\rm Impr. quark-gluon vert.}&$
- g\, \mfrac{\lambda^{a}_{\beta \alpha}}{2}\, \mfrac{r}{2}\,
\left[\, \sum_{\nu} \left( \,
\sigma_{\rho \nu}\, \sin ( (p-p')_{\nu} a )\, \right)\, 
\cos ( (p-p')_{\rho} \mfrac{a}{2} )\, \right]\, e^{-i p'_{\rho} 
\frac{a}{2}}\, e^{i p_{\rho}\frac{a}{2}}$\\[7pt]
\hline
\mbox{}&\mbox{}\\[-7pt]
\mbox{\rm Impr. quark-quark vert.}&$ - \mfrac{r}{4}\, \delta_{\alpha \beta}\,
\sum_{\mu}\, \gamma_{\mu}\, \left[\, e^{i p_{\mu} a}\, -\,
e^{ - i p_{\mu} a}\, \right]$\\[7pt]
\hline
\mbox{}&\mbox{}\\[-10pt]
\mbox{\rm quark-gluon-quark vert.}&$ (- i\,  g\, a) \mfrac{r}{4}\,
\mfrac{\lambda^{a}_{\beta \alpha}}{2}\,
\gamma_{\rho}\, \left[\, e^{\pm i p_{\rho} a}\, +\,
e^{ \mp i p_{\rho} a}\, e^{ - i q_{\rho} a}\,\right]$\\[7pt]
\hline
\hline
\end{tabular}
\caption{Feynman rules for the lattice
Wilson and SW--Clover actions for light quarks.}
\label{tabfeynruleslight}
\end{table}

To be definite, let us consider a generic $\Delta B=2$ effective operator
with arbitrary $\Gamma_{1}$ and $\Gamma_{2}$ dirac matrices
\be
\tilde{O}(n)\, =\,\left( h^{\dagger}(n)\, \Gamma_{1}\, \psi(n)\, \right)\,
\left( \tilde{h}(n)\, \Gamma_{2}\, \psi(n)\, \right)
\ee
The improved operator $\tilde{O}^{I}$ is obtained by rotating the light--quark
fields according to eq.(\ref{psirot}), keeping terms of order $O(a^{2})$
for the second method of improvement \cite{frezzoti},
\begin{eqnarray} \label{imprope}
\tilde{O}^{I}(n) &=& \tilde{O}(n)\, -\, a\, \mfrac{r}{2}\,
\left[\, \left( h^{\dagger}(n)\, \Gamma_{1}\,
\vec{\ddir}\, \psi(n)\, \right)\,
\left( \tilde{h}(n)\, \Gamma_{2}\, \psi(n)\, \right)\right.\nonumber\\
&+& \left.\left( h^{\dagger}(n)\, \Gamma_{1}\, \psi(n)\, \right)\,
\left( \tilde{h}(n)\, \Gamma_{2}\, \vec{\ddir}\,\psi(n)\, \right)\right]\nonumber\\
&+& a^{2}\, \left(\mfrac{r}{2}\right)^{2}\,
\left( h^{\dagger}(n)\, \Gamma_{1}\,
\vec{\ddir}\, \psi(n)\, \right)\,
\left( \tilde{h}(n)\, \Gamma_{2}\, \vec{\ddir}\, \psi(n)\, \right)
\end{eqnarray}
Similarly, we can improve the $\Delta B = 0$ operators.

From eqs.(\ref{wilson}), (\ref{clover}) and (\ref{imprope}), we can obtain the Feynman rules which
are given in Table \ref{tabfeynruleslight}.
In this Table, the propagator 
corresponds to a quark of momentum $p$ and bare mass $m_{0}$, 
and color conservation is
understood.  Both the Wilson and the improved quark--gluon vertex, the latter 
coming from the SW term in action (\ref{clover}),  
correspond to an incoming quark of color $\alpha$ and momentum $p$,
outgoing quark of color $\beta$ and momentum $p'$ and incoming gluon 
of color index $a$ and Lorentz index $\rho$. The other two vertices in Table 2
come from the rotation of the four--quark operators in eq.(\ref{imprope}). In
the improved quark--quark vertex we have an incoming or outgoing
light--quark of color $\alpha$ and momentum $p$.
In the quark--gluon--quark vertex,
we have in addition an incoming gluon 
with momentum $q$, color index $a$ and Lorentz index $\rho$ and
the sign of the momentum $p$ in the exponentials is $+$ for an incoming 
light quark and $-$ for an outgoing one.
For the gluon, we use the standard Wilson action \cite{wilson}. 

\section{The strategy: the subtle point of the renormalization scheme
dependence.}
\label{strategy}

Our aim is to express the continuum HQET operators at the scale 
$\mu=a^{-1}$ as a linear combination of lattice 
HQET operators at this scale in such a way that all continuum matrix elements
be equal to their corresponding lattice counterparts up to one--loop order in 
perturbation theory.

The procedure is well known. The relation between the
continuum and lattice operators to one--loop can be written as
\be \label{relconlat}
O_{i}^{con}\, =\, \sum_{j}\, [\, \delta_{ij}\, \,+\, \mfrac{g^{2}}{16\, \pi^{2}}\,
Z_{i j}\, ]\, O^{lat}_{j}
\ee
By sandwiching eq.(\ref{relconlat}) between appropriate, but arbitrary,
initial and final states, one can obtain the renormalization constants $Z$.
In fact, some amplitudes are evaluated up to one--loop in perturbation
theory both in the continuum and the lattice theory. After renormalization one
gets
\be
\langle O_{i}^{con(lat)}\rangle\, =\, 
\sum_{j}\, [\, \delta_{ij}\, \,+\, \mfrac{g^{2}}{16\, \pi^{2}}\,
c^{con(lat)}_{i j}\, ]\, \langle O^{con(lat)}_{j} \rangle^{(0)}
\ee
where $(0)$ denotes the bare matrix element. Thus, the constants
$Z$ are obtained by simply subtracting at the scale given,
$Z_{i j}\, =\, c^{con}_{i j}\, -\, c^{lat}_{i j}$; this is the matching
procedure.

Note that the amplitudes can be regularized and renormalized using, in general, 
different schemes in both theories. 
Some subtle points to be stressed are the following:
\begin{enumerate}
\item the matching must be calculated with the same 
action as for the numerical simulation because the matching coefficients
depend on the action to be simulated.
\item due to the breaking of chiral symmetry induced by the Wilson term for 
light quarks, the original operators can mix with lattice operators with
different chirality.
\item the matching constants $Z$'s depend on the renormalization procedure
chosen to define the operators in the continuum. This dependence will
cancel with the remaining scheme dependence of the running in step 2 of the
Introduction. This subtle point will be studied in detail in a forthcoming paper.
\end{enumerate}
Therefore, it is very important to clearly define the regularization and 
renormalization schemes we utilize in our calculation.
Many prescriptions can be chosen to regularize
the corresponding Feynman integrals: NDR (Na\"{\i}ve Dimensional Regularization),
DRED (Dimensional Reduction), HV (the 't Hooft--Veltman scheme) \ldots etc. 
Different regularization schemes will
give rise to different finite parts in the amplitudes and hence to
different matching coefficients. 
Having selected the regularization procedure, we have to choose the
renormalization scheme to eliminate ultraviolet divergences, for example
MOM (momentum subtraction), $\overline{MS}$ \ldots etc.  

Apparently, the continuum operators are now completely defined. A more
careful analysis demonstrates, however, that this is not the case. 
In fact, there is another prescription to be set, namely, how to deal 
with the evanescent operators which unavoidably are generated \cite{marco}. 
We will explicitly show that different treatments of the evanescent operators
will give rise to different finite contributions to the amplitude
even when we return to four dimensions. The mechanism is clear:
evanescent operators are operators which vanish in four dimensions,
in other words, they are operators of order $O(n-4)$, with $n=4+2\epsilon$.
When an evanescent operator is produced in a divergent diagram, terms of the 
form $1/\epsilon\, \times\, \mbox{\rm evanescent}$ arise. Therefore,
$O(\epsilon^{0})$ contributions are generated and their value depend
on the prescription we use to subtract the evanescent operators.

To sum, it is not sufficient to specify the scheme we use 
to renormalize our Feynman diagrams, we have also to define the
prescription to eliminate the evanescent operators \cite{marco}.
This fact results in the existence of families of $\overline{MS}$ schemes 
which members differ in the treatment applied to the evanescent
operators:  we can, for instance, simply subtract them together
with the divergent pole or project the amplitudes onto some convenient, but
arbitrary, operator basis using some well--defined 
projectors. Note that in the second case, the projectors can be
chosen in many different ways, each of which will define a different member
in the corresponding scheme family.
Of course, we can use our favourite prescription. It is immaterial
with which  one we perform the calculation. 
At the end, when the continuum QCD operator will be constructed in terms of HQET 
lattice operators, all dependence on the intermediate 
renormalization scheme used to define continuum HQET operators will drop and 
the final result will depend on the definition of the QCD operators only. 
The key point here is to be consistent during the whole procedure so that no
spurious scheme dependence creep into our calculation.
   
\section{The calculation.}

We compute the matrix elements of the $\Delta B=2$ operators in 
eq.(\ref{allopero}) between an initial $B^{0}(\bar{b}q)$ state and a 
$\bar{B}^{0}(\bar{q}b)$ final state. For the $\Delta B =0$ operators in eq.(\ref{alloperq})
we calculate the matrix elements between an initial and a final $\bar{B}^{0}(\bar{q}b)$ 
state neglecting penguin diagrams \cite{taubbs}.
We set the light--quark mass to zero and take our
momentum configuration to be vanishing incoming and outgoing momenta of both
light and heavy quarks.  To  eliminate the infrared divergences which appear
at zero external momenta, we give the gluon a mass, $\lambda$. This procedure
is justified at one loop order.

On the lattice, the Feynman diagrams to be evaluated are depicted in 
fig.~1. The box represents an insertion of the four--fermion operator.
The cross vertex stands for the contribution coming from the rotated part 
of the improved operator (see the entries for the improved quark-quark and
quark-gluon-quark vertices in Table 
\ref{tabfeynruleslight}). 
Note that each light quark--gluon vertices in all diagrams consists of
the Wilson vertex plus the improved contribution coming from the Clover
action (see entries for the Wilson and Improved quark-gluon vertices in
Table \ref{tabfeynruleslight}).
Ultraviolet divergences are regulated by the lattice cut--off, $a$. 

%Following ref.\cite{}, we have split the set of diagrams into three subsets:
%the subset A, containing the diagrams with no contributions from the rotated
%part of the improved operator; the subset B, which includes the
%diagrams with one improved quark--quark or quark--gluon--quark vertex
%coming from the rotated part of the operator and set C, containing
%diagrams with two improved quark--quark or quark--gluon--quark vertices.

In practice, we project the lattice amplitudes 
onto the Dirac basis ${\cal B}\, =\, \{ \gamma_{\mp}$, $\gamma_{\mp \pm}$,
$\gamma^{s}_{\mp}$, $\gamma^{s}_{\mp\pm}$, $\sigma_{\mp}\, \}$, where
$\gamma_{\mp}$ denotes the Dirac structure 
$\gamma^{\mu} (1\mp\gamma_{5}) \otimes  \gamma_{\mu} (1\mp\gamma_{5})$,
$\gamma_{\mp \pm}$ is 
$\gamma^{\mu} (1\mp\gamma_{5}) \otimes  \gamma_{\mu} (1\pm\gamma_{5})$,
$\gamma^{s}_{\mp}$ is
$(1\mp\gamma_{5}) \otimes  (1\mp\gamma_{5})$,
$\gamma^{s}_{\mp \pm}$ is
$(1\mp\gamma_{5}) \otimes  (1\pm\gamma_{5})$ and finally
$\sigma_{\mp}$ is  
$\sigma^{\mu \nu} (1\mp\gamma_{5}) \otimes  \sigma_{\mu \nu} (1\mp\gamma_{5})$.
%${\cal B}\, =\, \{ O_{LL(RR)}$, $O_{LR(RL)}$, 
%$O^{S}_{LL(RR)}$, $O^{S}_{LR(RL)}$, $O^{\sigma}_{LL(RR)} \}$ 
%for $\Delta B=2$ 
%operators and similarly for the $\Delta B=0$ ones. 
%All operators in the basis
%are defined in eqs.(\ref{allopero}) and (\ref{alloperq}) except for
%$O(Q)^{\sigma}_{LL(RR)}$ which have the Dirac structure $\sigma^{\mu \nu}\,
%(1\mp \gamma_{5})\otimes  \sigma_{\mu \nu}\,
%(1\mp \gamma_{5}) $. Notice also that $O(Q)^{\sigma}_{LR(RL)}$ vanishes in 
Notice also that $\sigma_{\mp \pm}$ vanishes in
four dimensions. The projectors are defined by (the sign is $+$ in Euclidean
space and $-$ in Minkowski space) 
\begin{eqnarray} \label{projectors}
%P_{vl(r)vl(r)} (\Gamma_{1} \otimes \Gamma_{2})&=& Tr\left[\, 
P_{\gamma_{\mp}} (\Gamma_{1} \otimes \Gamma_{2})&=& Tr\left[\, 
\Gamma_{1}\, \gamma^{\mu}\, (\, 1\, \pm\, \gamma_{5}\, )\, 
\Gamma_{2}\, \gamma_{\mu}\, (\, 1\, \pm\, \gamma_{5}\, )\,
\right]/(32\, n\, (2-n))\nonumber\\
%P_{vl(r)vr(l)} (\Gamma_{1} \otimes \Gamma_{2})&=& Tr\left[\, 
P_{\gamma_{\mp \pm}} (\Gamma_{1} \otimes \Gamma_{2})&=& Tr\left[\, 
\Gamma_{1}\, (\, 1\, \mp\, \gamma_{5}\, )\, 
\Gamma_{2}\, (\, 1\, \pm\, \gamma_{5}\, )\,
\right]/(32\, n)\nonumber\\
%P_{sl(r)sl(r)} (\Gamma_{1} \otimes \Gamma_{2})&=& Tr\left[\, 
P_{\gamma^{s}_{\mp}} (\Gamma_{1} \otimes \Gamma_{2})&=& Tr\left[\, 
\Gamma_{1}\, (\, 1\, \mp\, \gamma_{5}\, )\, 
\Gamma_{2}\, (\, 1\, \mp\, \gamma_{5}\, )\,
\right](-n^{2}+9 n-16)/(256\, (n-2))\nonumber\\
&\pm& Tr\left[\, 
\Gamma_{1}\, \sigma^{\mu \nu}\, (\, 1\, \mp\, \gamma_{5}\, )\, 
\Gamma_{2}\, \sigma_{\mu \nu}\, (\, 1\, \mp\, \gamma_{5}\, )\,
\right]/(256\, (2-n))\nonumber\\ 
%P_{sl(r)sr(l)}(\Gamma_{1} \otimes \Gamma_{2}) &=& Tr\left[\, 
P_{\gamma^{s}_{\mp \pm}}(\Gamma_{1} \otimes \Gamma_{2}) &=& Tr\left[\, 
\Gamma_{1}\, \gamma^{\mu}\, (\, 1\, \pm\, \gamma_{5}\, )\, 
\Gamma_{2}\, \gamma_{\mu}\, (\, 1\, \mp\, \gamma_{5}\, )\,
\right]/(32\, n)\nonumber\\
%P_{tl(r)tl(r)}(\Gamma_{1} \otimes \Gamma_{2}) &=& Tr\left[\, 
P_{\sigma_{\mp}}(\Gamma_{1} \otimes \Gamma_{2}) &=& Tr\left[\, 
\Gamma_{1}\, \sigma^{\mu \nu}\, (\, 1\, \mp\, \gamma_{5}\, )\, 
\Gamma_{2}\, \sigma_{\mu \nu}\, (\, 1\, \mp\, \gamma_{5}\, )\,
\right]/(256\, n\, (n-1)\, (2-n))\nonumber\\
&\pm& Tr\left[\, 
\Gamma_{1}\, (\, 1\, \mp\, \gamma_{5}\, )\, 
\Gamma_{2}\, (\, 1\, \mp\, \gamma_{5}\, )\,
\right]/(256\, (2-n))
\end{eqnarray}
where $n=4$ ($n=4+2\epsilon$) for a four ($n$) dimensional gamma algebra and 
$\Gamma_{1,2}$ are arbitrary Dirac matrices. On the lattice, $\gamma_{\mu}$ are the Euclidean gamma matrices 
satisfying the anticommutation relation
$\left\{\gamma_{\mu}, \gamma_{\nu}\right\}=2\, \delta_{\mu \nu}$ in four dimensions 
and $\sigma_{\mu \nu} = 1/2 \left[\gamma_{\mu}, \gamma_{\nu}\right]$.
%In eq.(\ref{projectors}),
%$vl(r)vl(r)$ denotes the Dirac structure 
%$\gamma^{\mu} (1\mp\gamma_{5}) \otimes  \gamma_{\mu} (1\mp\gamma_{5})$,
%$vl(r)vr(l)$ is 
%$\gamma^{\mu} (1\mp\gamma_{5}) \otimes  \gamma_{\mu} (1\pm\gamma_{5})$,
%$sl(r)sl(r)$ is
%$(1\mp\gamma_{5}) \otimes  (1\mp\gamma_{5})$,
%$sl(r)sr(l)$ is
%$(1\mp\gamma_{5}) \otimes  (1\pm\gamma_{5})$ and finally
%$tl(r)tl(r)$ is  
%$\sigma^{\mu \nu} (1\mp\gamma_{5}) \otimes  \sigma_{\mu \nu} (1\mp\gamma_{5})$.
With our definition, all operators in the basis ${\cal B}$ project
back onto themselves.

Since HQET fields $h$ satisfy 
$\gamma_{0}\,  u_{h} = u_{h}$ and  $\gamma_{0}\, v_{h} = - v_{h}$, where 
$u_{h}$ and $v_{h}$ are the 
spinors for a heavy quark and antiquark, we can reduce the number of
independent lattice amplitudes through the following Euclidean--space 
relationships
\begin{eqnarray} \label{reduamplhqet} 
\lefteqn{\left[\bar{u}_{h}\, \sigma^{\mu \nu}\, (1\, \mp\, \gamma_{5})\, v_{q}\, 
\bar{v}_{h}\, \sigma_{\mu \nu}\, (1\, \mp\, \gamma_{5})\, u_{q} \right]\,
=}&&\nonumber\\
&&- 4\, \left( 
\left[\bar{u}_{h}\, (1\, \mp\, \gamma_{5})\, v_{q}\, 
\bar{v}_{h}\, (1\, \mp\, \gamma_{5})\, u_{q} \right]\,
+\, \left[\bar{u}_{h}\,\gamma^{\mu}\, (1\, \mp\, \gamma_{5})\, v_{q}\,
\bar{v}_{h}\,  
 \gamma_{\mu}\, (1\, \mp\, \gamma_{5})\, u_{q} \right]\right)\hfill\nonumber\\
\lefteqn{\left[\bar{u}_{h}\, \sigma^{\mu \nu}\, (1\, \mp\, \gamma_{5})\, v_{q}\, 
\bar{v}_{q}\,   
 \sigma_{\mu \nu}\, (1\, \mp\, \gamma_{5})\, u_{h} \right]\, =}&& \\
&&+ 4\, \left( 
\left[\bar{u}_{h}\, (1\, \mp\, \gamma_{5})\, v_{q}\, 
\bar{v}_{q}\, (1\, \mp\, \gamma_{5})\, u_{h} \right]\,
-\, \left[\bar{u}_{h}\,\gamma^{\mu}\, (1\, \mp\, \gamma_{5})\, v_{q}\,
\bar{v}_{q}\,  
 \gamma_{\mu}\, (1\, \pm\, \gamma_{5})\, u_{h} \right]\right)\nonumber
\end{eqnarray}
where $u_{q}$ and $v_{q}$ are the light quark and antiquark spinors.
The first relation can be used to reduce $\Delta B=2$ amplitudes and the
second for $\Delta B=0$ ones.

Moreover, since Fierz transformations are well
defined in four dimensions, we have for $\Delta B =2$ operators
\begin{eqnarray} \label{reduampl}
{\cal F}\left[\gamma^{\mu}\, (1\, \mp\, \gamma_{5})\, \otimes\,  
 \gamma_{\mu}\, (1\, \mp\, \gamma_{5}) \right]  &=& -\,
\left[\gamma^{\mu}\, (1\, \mp\, \gamma_{5})\, \otimes\,  
 \gamma_{\mu}\, (1\, \mp\, \gamma_{5}) \right]\nonumber\\
{\cal F}\left[\gamma^{\mu}\, (1\, \mp\, \gamma_{5})\, \otimes\,  
 \gamma_{\mu}\, (1\, \pm\, \gamma_{5}) \right]  &=& 2\,
\left[(1\, \pm\, \gamma_{5})\, \otimes\,  
 (1\, \mp\, \gamma_{5}) \right]\nonumber\\
{\cal F}\left[(1\, \mp\, \gamma_{5})\, \otimes\,  
 (1\, \mp\, \gamma_{5}) \right]  &=&
\left[(1\, \mp\, \gamma_{5})\, \otimes\,  
 (1\, \mp\, \gamma_{5}) \right]\nonumber\\
&+& \mfrac{1}{2}\, \left[\gamma^{\mu}\, (1\, \mp\, \gamma_{5})\, \otimes\,  
 \gamma_{\mu}\, (1\, \mp\, \gamma_{5}) \right] 
\end{eqnarray}   
where ${\cal F}$ denotes the Fierz transformation and we have omitted for
simplicity the external spinors.

In the continuum, only diagrams $D_{1}$ to $D_{6}$ contribute to the amplitude, for
neither improved nor tadpole vertices appear (see Table \ref{tabfeynruleshqet}).
Ultraviolet divergences will be regulated using dimensional regularization.
Unlike the lattice case, evanescent operators do
arise in the continuum so that we have to give a prescription to
deal with them.

If the gamma algebra is in $n$ dimensions, as in NDR, 
the relations in  eq.(\ref{reduampl}) are not valid because
Fierz transformations are not well defined.
Moreover, since the ''magic'' formulae for 
reducing the product of three gamma matrices is correct only in four
dimensions, eq.(\ref{reduamplhqet}) holds up to evanescent 
operator contributions. For instance, the continuation of the 
first relation in eq.(\ref{reduamplhqet}) to $n$ dimensions can be defined by
\begin{eqnarray} \label{magicrel}
\lefteqn{\left[\sigma^{\mu \nu}\, (1\, \mp\, \gamma_{5})\, \otimes\,  
 \sigma_{\mu \nu}\, (1\, \mp\, \gamma_{5}) \right]  =}\nonumber\\
&& f(n)\, \left( 
\left[\gamma^{\mu}\, (1\, \mp\, \gamma_{5})\, \otimes\,  
 \gamma_{\mu}\, (1\, \mp\, \gamma_{5}) \right]\, +\,
 \left[(1\, \mp\, \gamma_{5})\, \otimes\, 
(1\, \mp\, \gamma_{5}) \right]\right)\, +\, \sigma_{f}
\end{eqnarray}
where $f(n)$ depends on the choice of the evanescent operator $\sigma_{f}$. 
A na\"{\i}ve prescription is to impose that
$f(n)=4$ as in four dimensions (notice that there is a change of sign between
Minkowski and Euclidean spaces). However, we can equally well 
take $f(n)=n$ by redefining $\sigma_{f}$. A very interesting choice is $f(n)=4 (1+1/6\, \epsilon)$.
This prescription is useful because leads to
results which are Fierz symmetric.
Notice that all three prescriptions are valid and 
differ only in the definition of the evanescent operator $\sigma_{f}$.

If we use DRED, where the positions and momenta are continued to $n$ 
dimensions whereas all other tensors, in particular the gamma algebra,
are leaving in four dimensions \cite{siegel}, 
eqs.(\ref{reduamplhqet}) and  (\ref{reduampl})
hold but new independent operators arise \cite{martinelli} 
\begin{eqnarray}\label{drednewope}
\bar{\gamma}_{\mp} &\equiv& \left[\gamma^{\mu}\, (1\, \mp\, \gamma_{5})\, \otimes\,  
\gamma^{\nu}\, (1\, \mp\, \gamma_{5}) \right]\, g^{(n)}_{\mu \nu}\, \equiv\,
\mfrac{n}{4}\, \gamma_{\mp}\, +\,  \gamma^{f}_{\mp}\nonumber\\
\bar{\gamma}_{\mp \pm} &\equiv& \left[\gamma^{\mu}\, (1\, \mp\, \gamma_{5})\, \otimes\,  
\gamma^{\nu}\, (1\, \pm\, \gamma_{5}) \right]\, g^{(n)}_{\mu \nu}\, \equiv\,
\mfrac{n}{4}\, \gamma_{\mp \pm}\, +\, \gamma^{f}_{\mp \pm}\nonumber\\
\bar{\sigma}_{\mp} &\equiv& \left[\sigma^{\mu \nu}\, (1\, \mp\, \gamma_{5})\, \otimes\,  
\sigma^{\rho \tau}\, (1\, \mp\, \gamma_{5}) \right]\, g^{(4)}_{\mu \rho}\,
g^{(n)}_{\nu \tau}\, \equiv\, \mfrac{n}{4}\, \sigma_{\pm}\, +\, \sigma^{f}_{\mp}
\end{eqnarray}
where $g^{(4)}_{\mu \nu}$ and $g^{(n)}_{\mu \nu}$ are the metric tensors in four and $n$ dimensions
respectively. The Dirac structures $\gamma_{\mp}$, $\gamma_{\mp\pm}$ and
$\sigma_{\mp}$ are defined similarly to those with a bar except for the fact
that Lorentz indices are contracted in four dimensions only.
We have also defined three evanescent operators 
$\gamma^{f}_{\mp}$, $\gamma^{f}_{\mp \pm}$ and $\sigma^{f}_{\mp}$ which will
prove to be useful in the calculation.
Notice that the operator $\bar{\sigma}_{\pm \mp}$, with the same notation as
in eq.(\ref{drednewope}), is an evanescent operator because $\sigma_{\pm \mp}$
vanishes in four dimensions.

In the scheme proposed by 't Hooft and Veltman (HV) \cite{thooft}, $\gamma_{5}$
anticommutes with the gamma matrices in four dimensions but commutes with
those in $n-4$ dimensions. In addition, the rule to consistently 
define the coupling to chiral fields in the Standard Model implies that
Dirac matrices in the four--fermion operators must be in four dimensions. 

In order to show the dependence of the matching on the prescription
chosen to define the continuum operators, we will perform the calculation
in several schemes, namely,
\begin{description}
\item[\bf NDR$_{\xi}$:] NDR--$\overline{MS}$, where we project onto the
basis ${\cal B}$ in $n$ dimensions with eq.(\ref{projectors}). The HQET magic relation 
for $\Delta B=2$ operators (\ref{magicrel}), used to reduce the number of
independent amplitudes, is taken with $f(n)=4\, (\, 1\, +\, \xi\, \epsilon\, )$,
where $\xi$ is an arbitrary real number. A similar prescription is used for
$\Delta B=0$ operators.
\item[\bf DRED I:] DRED--$\overline{MS}$, where we subtract with the basis 
${\cal B}$ plus the evanescent operators $\{\, 
\gamma_{\mp}\, -\, \bar{\gamma}_{\mp}$, 
$\gamma_{\mp \pm}\, -\, \bar{\gamma}_{\mp \pm}$, 
$\sigma_{\mp}\, -\, \bar{\sigma}_{\mp}$, 
$\bar{\sigma}_{\mp \pm}\, \}$.  
\item[\bf DRED II:]  DRED--$\overline{MS}$, where we subtract with the basis 
${\cal B}$ plus the evanescent operators $\{\, \gamma^{f}_{\mp}$,  
$\gamma^{f}_{\mp \pm}$, $\sigma^{f}_{\mp}$, $\bar{\sigma}_{\mp \pm}\, \}$. 
\item[\bf HV:] HV--$\overline{MS}$, where we project onto the basis ${\cal B}$
in four dimensions with eq.(\ref{projectors}).
\end{description}
Notice that DRED II is equivalent to eliminate 
the evanescent operators by simply projecting them out 
because $P_{i}(\gamma_{f})=0$ ($\gamma_{f}$ being any of the 
evanescent operators in the basis)
for all projectors $P_{i}$ in four dimensions in eq.(\ref{projectors}).

\section{The Results.}

The continuum HQET operators $\tilde{O}_{LL}$, $\tilde{O}_{LR}$,
$\tilde{O}^{S}_{LL}$ and $\tilde{O}^{S}_{LR}$ can be written in terms of lattice operators as
\begin{eqnarray} \label{eqocon}
\tilde{O}_{LL}(\mu) &=& \left( 1\, +\, \mfrac{g_{s}^{2}}{16\, \pi^{2}}\,
\left[\, -4\, \ln\! \left(\mfrac{\lambda^{2}}{\mu^{2}}\right)\, +\, \mfrac{4\, +\, 3\, 
z\, +\, 3\, z'''}{3} \right]\right.\nonumber\\ 
&+& \left. \mfrac{g_{lat}^{2}}{16\, \pi^{2}}\, \left[ 4\, \ln (\lambda^{2} a^{2})\, +\, (D_{LL}\, +\,
D^{I}_{LL}\, +\, D^{II}_{LL}) \right]\, \right)\, O^{lat}_{LL}\nonumber\\
&+& \mfrac{g_{lat}^{2}}{16\, \pi^{2}}\,( D_{RR}\, +\, D^{I}_{RR}\, +\, D^{II}_{RR})\, 
O_{RR}^{lat}\nonumber\\
&+& \mfrac{g_{lat}^{2}}{16\, \pi^{2}}\, ( D_{N}\, +\, D^{I}_{N} )\, O^{lat}_{N}\\
\tilde{O}_{LR}(\mu) &=& \left( 1\, +\, \mfrac{g_{s}^{2}}{16\, \pi^{2}}\,
\left[\, -7/2\, \ln\! \left(\mfrac{\lambda^{2}}{\mu^{2}}\right)\, +\, \mfrac{23\, +\, 18\, z\, \, -\, 
z'\, +\, 14\, z'''}{12} \right]\right.\nonumber\\
&+&\, \left.\mfrac{g_{lat}^{2}}{16\, \pi^{2}}\, \left[ 7/2\, \ln (\lambda^{2} a^{2})\, +\, (D_{LR}\, +\,
D^{I}_{LR}\, +\, D^{II}_{LR}) \right]\, \right)\, O^{lat}_{LR}\nonumber\\
&+& \left(\mfrac{g_{s}^{2}}{16\, \pi^{2}}\,\left[ 3\, \ln\! \left(\mfrac{\lambda^{2}}{\mu^{2}}\right)\,
+\, \mfrac{-1\, +\, 2\, z\, -\, z'\, -\, 2\, z'''}{2} \right]\right.\nonumber\\
&+&\, \left.\mfrac{g_{lat}^{2}}{16\, \pi^{2}}\, \left[\, -3\, \ln (\lambda^{2} a^{2})\, +\,
( \bar{D}^{S}_{RL}\, +\, \bar{D}^{S\, I}_{RL}\, +\, \bar{D}^{S\, II}_{RL})\, \right]\,\right)\,
O^{lat\: S}_{RL}\nonumber\\
&+& \mfrac{g_{lat}^{2}}{16\, \pi^{2}}\,
( D_{M}\, +\, D^{I}_{M} )\, O^{lat}_{M}\\
\label{eqocon3}
\tilde{O}^{S}_{LL}(\mu) &=& \left( 1\, +\, \mfrac{g_{s}^{2}}{16\, \pi^{2}}\,
\left[\, -4/3\, \ln\! \left(\mfrac{\lambda^{2}}{\mu^{2}}\right)\, +\, 
\mfrac{1\, +\, 4\, z\, +\, z'\, +\, z''\, +\, 3\, z'''}{3} \right]\right.\nonumber\\
&+& \left.\, \mfrac{g_{lat}^{2}}{16\, \pi^{2}}\,\left[ 4/3\, \ln (\lambda^{2} a^{2})\, +\, (D^{S}_{LL}\, +\,
D^{S\, I}_{LL}\, +\, D^{S\, II}_{LL}) \right]\, \right)\, O^{lat\: S}_{LL}\nonumber\\
&+& \left(\mfrac{g_{s}^{2}}{16\, \pi^{2}}\,\left[ 2/3\, \ln\! \left(\mfrac{\lambda^{2}}{\mu^{2}}\right)\,
-\, \mfrac{3\, +\, z'\, +\, z''\, +\, 3\, z'''}{24} \right]\right.\nonumber\\ 
&+& \left.\, \mfrac{g_{lat}^{2}}{16\, \pi^{2}}\,\left[ -2/3\, \ln (\lambda^{2} a^{2})\, +\, (\bar{D}_{LL}\, +\,
\bar{D}^{I}_{LL}\, +\, \bar{D}^{II}_{LL}) \right]\, \right)\, O^{lat}_{LL}\nonumber\\
&+& \mfrac{g_{lat}^{2}}{16\, \pi^{2}}\,
( \bar{D}_{RR}\, +\, \bar{D}^{I}_{RR}\, +\, \bar{D}^{II}_{RR})\, O^{lat}_{RR}\nonumber\\
&+& \mfrac{g_{lat}^{2}}{16\, \pi^{2}}\,
( D_{P}\, +\, D^{I}_{P} )\, O^{lat}_{P}\\
\tilde{O}^{S}_{LR}(\mu) &=& \left( 1\, +\, \mfrac{g_{s}^{2}}{16\, \pi^{2}}\,
\left[ \, -7/2\, \ln\! \left(\mfrac{\lambda^{2}}{\mu^{2}}\right)\, +\, 
\mfrac{22\, +\, 19\, z\, +\, 15\, z'''}{12}\right]\right.\nonumber\\
&+& \left.\, \mfrac{g_{lat}^{2}}{16\, \pi^{2} }\, +\, \left[ 7/2\, \ln (\lambda^{2} a^{2})\, +\, (D^{S}_{LR}\, +\,
D^{S\, I}_{LR}\, +\, D^{S\, II}_{LR}) \right]\, \right)\, O^{lat\: S}_{LR}\nonumber\\
&+& \left(\mfrac{g_{s}^{2}}{16\, \pi^{2}}\,\left[ 3/4\, \ln\! \left(\mfrac{\lambda^{2}}{\mu^{2}}\right)\, 
+\, \mfrac{-\, 2\, +\, 3\, z\, -\, z'''}{8}\right]\right.\nonumber\\
&+& \left.\, \mfrac{g_{lat}^{2}}{16\, \pi^{2}}\, \left[\, -3/4\, \ln (\lambda^{2} a^{2})\, +\,
( \bar{D}_{RL}\, +\, \bar{D}^{I}_{RL}\, +\, \bar{D}^{II}_{RL})\, \right]\, \right)\, O^{lat}_{RL}\nonumber\\
&+& \mfrac{g_{lat}^{2}}{16\, \pi^{2}}\,
( D_{Q}\, +\, D^{I}_{Q} )\, O^{lat}_{Q} \label{eqocon4}
\end{eqnarray}
where the analytical expressions for the lattice constants, $D$, can be found in
the appendix A. Their numerical values for the Wilson parameter $r=1.0$
are given in Table \ref{numdo}. In obtaining these numbers, we have used
a reduced value of the heavy--quark self--energy constant $e=4.53$ (to be
compared with the non reduced one $e=24.48$) which is consistent with 
fitting the time dependence of the Monte Carlo data through $A\, e^{B\, t}$.
Notice that we have explicitly separated the
contributions coming from the Wilson action (denoted by $D_{LL}$ \ldots etc),
the Clover action with improved operators not including $O(a^{2})$
terms (denoted by $D^{I}_{LL}$ \ldots etc) and including $O(a^{2})$
corrections (denoted by $D^{II}_{LL}$ \ldots etc).
The new lattice operators $O_{N}$, $O_{M}$, $O_{P}$ and $O_{Q}$ 
are defined by
\begin{eqnarray} \label{defnmpq}
O^{lat}_{N} & = & O^{lat}_{LR}\, +\, O^{lat}_{RL}\, +\, 2\, (\,
O^{lat\, S}_{LR}\, +\, O^{lat\, S}_{RL}\, )\nonumber\\
O^{lat}_{M} & = & 3/2\, (\, O^{lat}_{LL}\, +\, O^{lat}_{RR}\, )\,
+\, 4\, (\, O^{lat\, S}_{LL}\, +\, O^{lat\, S}_{RR}\, )\nonumber\\
O^{lat}_{P} & = & O^{lat}_{LR}\, +\, O^{lat}_{RL}\, +\, 6\, (\,
O^{lat\, S}_{LR}\, +\, O^{lat\, S}_{RL}\, )\nonumber\\
O^{lat}_{Q} & = & O^{lat}_{LL}\, +\, O^{lat}_{RR}\, +\, 8\, (\,
O^{lat\, S}_{LL}\, +\, O^{lat\, S}_{RR}\, )
\end{eqnarray}

%Numerical value of the constants
\begin{table}[!t] 
\begin{center}
\begin{tabular}{||l||r|r|r||l||r|r|r||}
\hline
\hline
\mbox{\rm Constant}&\mbox{\rm Wilson}&\mbox{\rm Impr.~I}&\mbox{\rm Impr.~II}&
\mbox{\rm Constant}&\mbox{\rm Wilson}&\mbox{\rm Impr.~I}&\mbox{\rm Impr.~II}\nonumber\\
\hline
$D_{LL}$&$-41.24$&$16.30$&$0.20$&$D^{S}_{LL}  $&$-30.87 $&$16.50 $&$0.04 $\nonumber\\
$D_{RR}$&$-1.60$&$-0.40 $&$-1.23$&$\bar{D}_{LL}$&$2.59 $&$0.05 $&$-0.04 $\nonumber\\
$D_{N} $&$-14.44$&$0.62$&$0 $&$\bar{D}_{RR}$&$0.40$&$0.10 $&$0.31 $\nonumber\\
$\mbox{}$&$\mbox{}$&$\mbox{}$&$\mbox{}$&$D_{P} $&$1.81 $&$-0.08 $&$0$\nonumber\\
\hline
$D_{LR}    $&$-37.83 $&$13.33 $&$-0.94 $&$D^{S}_{LR}$&$-37.83 $&$13.33 $&$-0.94 $  \nonumber\\
$\bar{D}^{S}_{RL}$&$9.80$&$1.10 $&$-0.73 $&$\bar{D}_{RL} $&$2.45$&$0.28 $&$-0.18 $  \nonumber\\
$D_{M}     $&$-7.22$&$0.31$&$0 $&$D_{Q}     $&$1.81 $&$-0.08 $&$0 $  \nonumber\\
\hline
\hline
\end{tabular}
\caption{Numerical values of the constants D for the Wilson parameter $r=1.0$.}
\label{numdo}
\end{center}
\end{table}

As can be seen from eqs.(\ref{eqocon}) to (\ref{eqocon4}), the lattice 
transcriptions of the continuum operators depend on the renormalization 
scheme used to define the latter, including the prescription to deal 
with the evanescent operators. 
If fact, $z$  to $z'''$ parameterize the scheme dependence in the continuum 
and are defined in Table \ref{zzz}.
A interesting particular value of $\xi$ is $\xi=1/3$ which corresponds to a NDR scheme which 
preserves Fierz symmetry, i.e.~renormalization and Fierz transformations 
commutate in the continuum scheme NDR$_{1/3}$ at one--loop order.
  
Notice that the results in eqs.(\ref{eqocon}) to (\ref{eqocon4}) are expressed in
terms of two coupling constants, namely, the continuum, $g_{s}$,
and the lattice, $g_{lat}$, ones. At one loop, however, we can consistently 
identify these two coupling constants because the difference is
$O(\alpha_{s}^{2})$. In this case, the logarithms combine giving
$\log(a^{2}\, \mu^{2})$ so that the infrared regulator 
$\lambda$ disappears, as it should be.  
Several recipes for improving the convergence of the lattice perturbative 
series by choosing a convenient 
perturbative expansion parameter have been proposed \cite{boosted}.
The choice of the optimal coupling constant is discussed for the
particular case of the operator $O_{LL}$ in the next section.

The lattice counterpart of the operator $\tilde{O}_{LL}$ is already known.
It was determined by Flynn {\it et al} \cite{flynn} for the Wilson action 
and by Borrelli and Pittori \cite{bp} for the SW-Clover action. We have
extended the calculation to all the HQET operators in
eq.(\ref{allopero}). Our results for $\tilde{O}_{LL}$ agree with 
ref.\cite{flynn} but disagree with ref.\cite{bp} in the 
sign of the contribution proportional to $O^{latt}_{RR}$ coming from 
the Feynman diagram $D_{6}$ in fig.~1. This error propagates to the constant
$D^{I}_{RR}$ which correct value is given in Table \ref{numdo} (to be compared
with the value quoted in ref.\cite{bp}, $D^{I}_{RR}=-2.58$.)

\begin{table}[!t] 
\begin{center}
\begin{tabular}{||l||c|c|c|c||}
\hline
\hline
\mbox{\rm Scheme}&$z$&$z'$&$z''$&$z'''$\nonumber\\
\hline
\mbox{\rm NDR$_{\xi}$}&$1$&$0$&$\xi$&$0$\nonumber\\
\mbox{\rm DRED I}&$0$&$0$&$0$&$0$\nonumber\\
\mbox{\rm DRED II}&$0$&$1$&$0$&$0$\nonumber\\
\mbox{\rm HV}&$0$&$0$&$0$&$1$\nonumber\\
\hline
\hline
\end{tabular}
\caption{Values of the scheme dependent parameters $z$'s.}
\label{zzz}
\end{center}
\end{table}
 
The operators with a $t^{a}$ insertion, denoted by $\tilde{Ot}_{i}$, 
can be readily obtained from the corresponding 
expressions in eqs.(\ref{eqocon}) to (\ref{eqocon4}) by simply replacing the
lattice operators $\tilde{O}_{i}$ with the operators $\tilde{Ot}_{i}$ and 
the operators $O_{N}$, $O_{M}$, $O_{P}$  and
$O_{Q}$, defined in eq.(\ref{defnmpq}), with $O_{N'}$, $O_{M'}$, $O_{P'}$ and
$O_{Q'}$ given by
\begin{eqnarray}
O^{lat}_{N'} & = & - \mfrac{1}{2}\, (\, O^{lat}_{LR}\, +\, O^{lat}_{RL}\, )\, -\, 
(\, O^{lat\, S}_{LR}\, +\, O^{lat\, S}_{RL}\, )\nonumber\\
O^{lat}_{M'} & = & -3/2\, (\, O^{lat}_{LL}\, +\, O^{lat}_{RR}\, )\,
-\, 2\, (\, O^{lat\, S}_{LL}\, +\, O^{lat\, S}_{RR}\, )\nonumber\\
O^{lat}_{P'} & = &  O^{lat}_{LR}\, +\, O^{lat}_{RL}\, -\, 6\, (\,
O^{lat\, S}_{LR}\, +\, O^{lat\, S}_{RL}\, )\nonumber\\
O^{lat}_{Q'} & = &  O^{lat}_{LL}\, +\, O^{lat}_{RR}\, -\, 4\, (\, 
O^{lat\, S}_{LL}\, +\, O^{lat\, S}_{RR}\, ).
\end{eqnarray}

For the continuum HQET $\Delta B=0$ operators, the results we have obtained are the 
following
\begin{eqnarray} \label{eqqcon}
\tilde{Q}_{LL}(\mu) &=& \left( 1\, +\, \mfrac{g_{s}^{2}}{16\, \pi^{2}}\,
\left[\, -4\, \ln\! \left(\mfrac{\lambda^{2}}{\mu^{2}}\right)\, +\, 
\mfrac{6\, +\, 4\, z\, +\, 4\, z'''}{3} \right]\right.\nonumber\\ 
&+& \left. \mfrac{g_{lat}^{2}}{16\, \pi^{2}}\, \left[\, 4\, \ln (\lambda^{2} a^{2})\, +\, (E_{LL}\, +\,
E^{I}_{LL})\, \right]\, \right)\, Q^{lat}_{LL}\nonumber\\
&+& \left(\mfrac{g_{s}^{2}}{16\, \pi^{2}}\, \left[ 3\, \ln\! \left(\mfrac{\lambda^{2}}{\mu^{2}}\right)\,
 -\, \mfrac{1\, +\, 4\, z\, +\, z'}{2} \right]\right.\nonumber\\
&+& \left. \mfrac{g_{lat}^{2}}{16\, \pi^{2}}\, \left[ -3\, \ln (\lambda^{2} a^{2})\,
+\, ( Et_{LL}\, +\, Et^{I}_{LL}\, +\, Et^{II}_{LL})\right]\, \right)\, Qt_{LL}^{lat}\nonumber\\
&+& \mfrac{g_{lat}^{2}}{16\, \pi^{2}}\, [ \bar{Et}_{RL}^{S}\, +\, \bar{Et}^{S\, I}_{RL}\, +\,
\bar{Et}^{S\, II}_{RL}]\, Qt_{RL}^{lat\: S}\nonumber\\
&+& \mfrac{g_{lat}^{2}}{16\, \pi^{2}}\, ( E_{R}\, +\, E^{I}_{R} )\, Q^{lat}_{R}\nonumber\\
&+& \mfrac{g_{lat}^{2}}{16\, \pi^{2}}\, ( Et_{R}\, +\, Et^{I}_{R} )\, Qt^{lat}_{R}\\
\tilde{Q}_{LR}(\mu) &=& \left( 1\, +\, \mfrac{g_{s}^{2}}{16\, \pi^{2}}\,
\left[\, -4\, \ln\! \left(\mfrac{\lambda^{2}}{\mu^{2}}\right) +\, 
\mfrac{6\, +\, 4\, z\, +\, 4\, z'''}{3} \right]\right.\nonumber\\ 
&+&\, \left.\mfrac{g_{lat}^{2}}{16\, \pi^{2}}\, \left[\, 4\, \ln (\lambda^{2} a^{2})\, +\, (E_{LR}\, +\,
E^{I}_{LR})\, \right]\, \right)\, Q^{lat}_{LR}\nonumber\\
&+& \left(\mfrac{g_{s}^{2}}{16\, \pi^{2}}\, [2\, -\, 2\, z\, +\, 2\, z''']\,
+\, \mfrac{g_{lat}^{2}}{16\, \pi^{2}}\, [ Et_{LR}\, +\, Et^{I}_{LR}\, 
+\, Et^{II}_{LR}]\,\right)\, Qt^{lat}_{LR}\nonumber\\
&+& \mfrac{g_{lat}^{2}}{16\, \pi^{2}}\, [ Et_{RL}\, +\, Et^{I}_{RL}\, 
+\, Et^{II}_{RL})\, ]\, Qt^{lat}_{RL}\nonumber\\
&+& \mfrac{g_{lat}^{2}}{16\, \pi^{2}}\, ( E_{S}\, +\, E^{I}_{S} )\, Q^{lat}_{S}\nonumber\\
&+& \mfrac{g_{lat}^{2}}{16\, \pi^{2}}\, ( Et_{S}\, +\, Et^{I}_{S} )\, Qt^{lat}_{S}\\
\tilde{Q}^{S}_{LL}(\mu) &=& \left( 1\, +\, \mfrac{g_{s}^{2}}{16\, \pi^{2}}\,
\left[\, -4\, \ln\! \left(\mfrac{\lambda^{2}}{\mu^{2}}\right) +\, 
\mfrac{6\, +\, 4\, z\, +\, 4\, z'''}{3} \right]\right.\nonumber\\
&+& \left.\, \mfrac{g_{lat}^{2}}{16\, \pi^{2}}\,\left[\, 4\, \ln (\lambda^{2} a^{2})\, +\, (E^{S}_{LL}\, +\,
E^{S\, I}_{LL})\, \right]\, \right)\, Q^{lat\: S}_{LL}\nonumber\\
&+& \left(\mfrac{g_{s}^{2}}{16\, \pi^{2}}\,\left[ 4\, \ln\! \left(\mfrac{\lambda^{2}}{\mu^{2}}\right)\, 
+\, \mfrac{-\, 5\, +\, z'\, +\, z''\, +\, z'''}{2} \right]\right.\nonumber\\ 
&+& \left.\, \mfrac{g_{lat}^{2}}{16\, \pi^{2}}\,\left[ -4\, \ln (\lambda^{2} a^{2})\, +\, (Et^{S}_{LL}\, +\,
Et^{S\, I}_{LL}\, +\, Et^{S\, II}_{LL}) \right]\, \right)\, Qt^{lat\: S}_{LL}\nonumber\\
&+& \left(\mfrac{g_{s}^{2}}{16\, \pi^{2}}\,\left[\, -\, \ln\! \left(\mfrac{\lambda^{2}}{\mu^{2}}\right)\,
+\, \mfrac{3\, -\, z'\, -\, z''}{2} \right]\right.\nonumber\\
&+& \left.\, \mfrac{g_{lat}^{2}}{16\, \pi^{2}}\,\left[\, \ln (\lambda^{2} a^{2})\, +\, (\bar{E}t_{LR}\, +\,
\bar{E}t^{I}_{LR}\, +\, \bar{E}t^{II}_{LR}) \right]\, \right)\, Qt^{lat}_{LR}\nonumber\\
&+& \, \mfrac{g_{lat}^{2}}{16\, \pi^{2}}\, [\bar{E}t_{RL}\, +\, \bar{E}t^{I}_{RL}\, +\, \bar{E}t^{II}_{RL}]\,
Qt^{lat}_{RL}\nonumber\\
&+& \mfrac{g_{lat}^{2}}{16\, \pi^{2}}\,
( E_{T}\, +\, E^{I}_{T} )\, Q^{lat}_{T}\nonumber\\
&+& \mfrac{g_{lat}^{2}}{16\, \pi^{2}}\,
( Et_{T}\, +\, Et^{I}_{T} )\, Qt^{lat}_{T}\\
\tilde{Q}^{S}_{LR}(\mu) &=& \left( 1\, +\, \mfrac{g_{s}^{2}}{16\, \pi^{2}}\,
\left[\, -4\, \ln\! \left(\mfrac{\lambda^{2}}{\mu^{2}}\right) +\, 
\mfrac{6\, +\, 4\, z\, +\, 4\, z'''}{3} \right]\right.\nonumber\\ 
&+& \left. \mfrac{g_{lat}^{2}}{16\, \pi^{2}}\, \left[\, 4\, \ln (\lambda^{2} a^{2})\, +\, (E^{S}_{LR}\, +\,
E^{S\: I}_{LR})\, \right]\, \right)\, Q^{lat\: S}_{LR}\nonumber\\
&+& \left(\mfrac{g_{s}^{2}}{16\, \pi^{2}}\, \left[ 3\, \ln\! \left(\mfrac{\lambda^{2}}{\mu^{2}}\right)\,
 -\, \mfrac{2\, +\, 3\, z\, -\, z'''}{2} \right]\right.\nonumber\\
&+& \left. \mfrac{g_{lat}^{2}}{16\, \pi^{2}}\, \left[\, -3\, \ln (\lambda^{2} a^{2})\,
+\, ( Et^{S}_{LR}\, +\, Et^{S\, I}_{LR}\, +\, Et^{S\, II}_{LR})\right]\, \right)\, 
Qt_{LR}^{lat\: S}\nonumber\\
&+& \, \mfrac{g_{lat}^{2}}{16\, \pi^{2}}\, [ Et_{RR}\, +\, Et^{I}_{RR}\, +\, Et^{II}_{RR}]\, 
Qt_{RR}^{lat}\nonumber\\
&+& \mfrac{g_{lat}^{2}}{16\, \pi^{2}}\, ( E_{U}\, +\, E^{I}_{U} )\, Q^{lat}_{U}\nonumber\\
&+& \mfrac{g_{lat}^{2}}{16\, \pi^{2}}\, ( Et_{U}\, +\, Et^{I}_{U} )\, Qt^{lat}_{U}
\label{eqqcon4}
\end{eqnarray}
where the analytical expressions for the lattice constants $E$ can be found in
the appendix A. Their numerical values for the Wilson parameter
$r=1.0$ and a reduced heavy--quark self--energy,
are given in Table \ref{numdq}. Notice that, as before, we have explicitly 
separated the contributions coming from the Wilson action (denoted by $E_{LL}$ \ldots etc),
the Clover action with improved operators not including $O(a^{2})$
terms (denoted by $E^{I}_{LL}$ \ldots etc) and including $O(a^{2})$
corrections (denoted by $E^{II}_{LL}$ \ldots etc).
The new lattice operators $Q_{R}$, $Q_{S}$, $Q_{T}$ and $Q_{U}$ 
are defined by
\begin{eqnarray}\label{defqnew}
Q^{lat}_{R} & = & -\, Q^{lat}_{LR}\, -\, Q^{lat}_{RL}\, +\, 2\, (\, 
Q^{lat\, S}_{LL}\, +\, Q^{lat\, S}_{RR}\, )\nonumber\\
Q^{lat}_{S} & = & -\, Q^{lat}_{LL}\, -\, Q^{lat}_{RR}\, +\, 2\, (\, 
Q^{lat\, S}_{LR}\, +\, Q^{lat\, S}_{RL}\, )\nonumber\\
Q^{lat}_{T} & = & Q^{lat\, S}_{LR}\, +\, Q^{lat\, S}_{RL}\nonumber\\
Q^{lat}_{U} & = & Q^{lat\, S}_{LL}\, +\, Q^{lat\, S}_{RR}
\end{eqnarray}
The operators $Qt_{R,S,T,U}$, with a $t^{a}$ insertion, are defined as those in 
eq.(\ref{defqnew}) simply replacing $Q$ with $Qt$.

\begin{table} 
\begin{center}
\begin{tabular}{||l||r|r|r||l||r|r|r||}
\hline
\hline
%\multicolumn{4}{||c||}{r=1.0}&\multicolumn{4}{|c||}{r=0.5}\\
%\hline
\mbox{\rm Constant}&\mbox{\rm Wilson}&\mbox{\rm Impr.~I}&\mbox{\rm Impr.~II}&
\mbox{\rm Constant}&\mbox{\rm Wilson}&\mbox{\rm Impr.~I}&\mbox{\rm Impr.~II}\nonumber\\
\hline
$E_{LL}       $&$-38.40$&$13.41$&$0 $&$E^{S}_{LL}   $&$-38.40$&$13.41$&$0 $   \nonumber\\
$Et_{LL}      $&$10.60$&$1.30$&$-0.11$&$Et^{S}_{LL}  $&$11.29$&$4.62 $&$0.05 $   \nonumber\\
$\bar{Et}^{S}_{RL}  $&$4.80$&$1.19$&$3.69$&$\bar{Et}_{LR}$&$-0.69$&$-3.32 $&$-0.16 $    \nonumber\\
$E_{R}        $&$9.63 $&$-0.41 $&$0 $& $\bar{Et}_{RL}$&$1.20$&$0.30 $&$0.92 $  \nonumber\\
$Et_{R}       $&$7.22$&$-0.31$&$0 $&$E_{T}        $&$9.63 $&$-0.41 $&$0 $     \nonumber\\
$\mbox{}$&$\mbox{}$&$\mbox{}$&$\mbox{}$&$Et_{T}       $&$7.22$&$-0.31$&$0
$\nonumber\\
\hline
$E_{LR}       $&$-38.40$&$13.41$&$0 $&$E^{S}_{LR}   $&$-38.40$&$13.41$&$0 $  \nonumber\\
$Et_{LR}      $&$8.53$&$-8.67$&$-0.60$&$Et^{S}_{LR}  $&$10.60$&$1.30$&$-0.11$ \nonumber\\
$Et_{RL}      $&$4.80$&$1.19$&$3.69$&$Et_{RR}      $&$1.20$&$0.30 $&$0.92 $   \nonumber\\
$E_{S}        $&$9.63 $&$-0.41 $&$0 $&$E_{U}        $&$9.63 $&$-0.41 $&$0 $  \nonumber\\
$Et_{S}       $&$7.22$&$-0.31$&$0 $&$Et_{U}       $&$7.22$&$-0.31$&$0 $    \nonumber\\
\hline
\hline
\end{tabular}
\caption{Numerical values of the constants E for the
Wilson parameter $r=1.0$.}
\label{numdq}
\end{center} 
\end{table}
 
The expressions of the $\Delta B=0$ continuum HQET operators with a $t^{a}$ can be obtained
from eqs.(\ref{eqqcon}) to (\ref{eqqcon4}) through the following procedure.
Let $\tilde{Qt}$ be a generic continuum $\tilde{Q}$--operator with a $t^{a}$ insertion. In terms of
lattice operators, it can be written as 
\[
\tilde{Qt}\, =\, \sum_{i}\, Ft_{i}\, 
Qt^{lat}_{i}\, +\, \sum_{i}\, F_{i}\, Q^{lat}_{i}
\]
where $Ft_{i}$ and $F_{i}$ are related to the corresponding
constants of $\tilde{Q}$ as follows
\begin{eqnarray}
&&\left. \begin{array}{l}
Ft_{i}=E_{i} + 7/6\, Et_{i}\\
F_{i} =2/9\, Et_{i}
\end{array} \right\} \textrm{ if }   i\, \neq R,S,T,U\nonumber\\
\nonumber\\
&&\left. \begin{array}{l}
Ft_{i}= -1/2\, Et_{i}\\
F_{i} =1/6\, E_{i}
\end{array} \right\} \textrm{ if } i\, =R,S,T,U
\end{eqnarray}

As a byproduct of our calculation, we have also obtained the anomalous
dimension matrix, $\hat{\gamma}$, at one loop, $\hat{\gamma}^{(0)}$,
defined by 
\be
\hat{\gamma}\, =\, \left(\mfrac{g^{2}}{16 \pi^{2}}\, \right)\, \hat{\gamma}^{(0)}\, 
+\,  \left(\mfrac{g^{2}}{16 \pi^{2}}\, \right)^{2}\, \hat{\gamma}^{(1)}\, \cdots
\ee 
for both the $\Delta B=2$ and $\Delta B=0$ operators.
In the continuum dimensional regularization $\hat{\gamma}^{(0)}$ is simply twice 
the residue of the $1/\epsilon$ pole and on the lattice it is twice
the coefficient of the $\log(\lambda^{2}\, a^{2})$. Therefore,
we get for the $\Delta B=2$ operators,
\be
\label{eqo:dim0}
\hat{\gamma}^{(0)}\, =\, \left[ \begin{array}{cccc}
-3\, (N-\mfrac{1}{N})&0&0&0\\
(1+\mfrac{1}{N})&-3 N+4+\mfrac{7}{N}&0&0\\
0&0&-3 (N - \mfrac{2}{N})&6\\
0&0&3/2&-3 (N - \mfrac{2}{N})
\end{array} \right]
\ee
in the basis $\{ \tilde{O}_{LL}$, $\tilde{O}^{S}_{LL}$, $\tilde{O}_{LR}$, 
$\tilde{O}^{S}_{RL}\}$ and where $N$
is the number of colors.

For the $\Delta B=0$ operators, we obtain the submatrices
\be
\label{eqq1:dim0}
\hat{\gamma}^{(0)}\, =\, \left[ \begin{array}{cccc}
-3\, (N-\mfrac{1}{N})&6&0&0\\
\mfrac{3}{2}(1-\mfrac{1}{N^{2}})&-\mfrac{3}{N}&0&0\\
0&0&-3 (N - \mfrac{1}{N})&6\\
0&0&\mfrac{3}{2}(1-\mfrac{1}{N^{2}})&-\mfrac{3}{N}
\end{array} \right]
\ee
in the basis $\{ \tilde{Q}_{LL}$, $\tilde{Qt}_{LL}$, $\tilde{Q}^{S}_{LR}$, 
$\tilde{Qt}^{S}_{LR}\}$, and
\be
\label{eqq2:dim0}
\hat{\gamma}^{(0)}\, =\, \left[ \begin{array}{cccc}
-3 (N - \mfrac{1}{N})&0&0&0\\
0&-3 (N - \mfrac{1}{N})&0&0\\
0&-2&-3 (N - \mfrac{1}{N})&8\\
-\mfrac{1}{2}(1-\mfrac{1}{N^{2}})&-(N - \mfrac{2}{N})&
2\, (1-\mfrac{1}{N^{2}})&(N - \mfrac{5}{N})
\end{array} \right]
\ee
in the basis $\{ \tilde{Q}_{LR}$, $\tilde{Qt}_{LR}$, $\tilde{Q}^{S}_{LL}$, 
$\tilde{Qt}^{S}_{LL}\}$.

\section{Reanalysis of the $B^0$-$\bar{B}^{0}$ mixing}

As we said in the previous section, the value of the lattice constant 
$D^{I}_{RR}$ quoted in ref.\cite{bp}, is incorrect. This constant
enters the continuum--lattice matching of the QCD operator $O_{LL}$, which
matrix element between $B^{0}$ and $\bar{B}^{0}$ states determines the 
theoretical prediction of the $B^0$-$\bar{B}^{0}$ mixing through
the so-called B--parameter $B_{B}$ \cite{gm}. In this section, we re--calculate 
the relevant renormalization constants and 
give the correct value of the B--parameter.

The B--parameter $B_{B}$ is the ratio of the matrix element of the operator
$O_{LL}$ to its vacuum insertion approximation
\be
B_{B}(\mu)\, \equiv\, \mfrac{\langle \bar{B}^{0} \mid
O_{LL}(\mu)\, \mid B^{0} \rangle}{\mfrac{8}{3}\, \mid\langle 0 \mid
A_{0} \mid B^{0} \rangle \mid^{2}}
\ee
The continuum QCD--lattice HQET matching for the 
operator $O_{LL}$ and the heavy--light axial current $A_{\mu}=\bar{b}\,
\gamma_{\mu}\, \gamma_{5}\, q$ can be written 
as
\begin{eqnarray}
O_{LL} &=& Z_{O_{L}}\, O_{LL}^{lat}\, +\, Z_{O_{R}}\, 
O_{RR}^{lat}\, +\, Z_{O_{N}}\, O_{N}^{lat}\, +\, Z_{O_{S}}\, O^{lat\:
S}_{LL}\nonumber\\
A_{\mu} &=& Z_{A}\, A_{\mu}^{lat} 
\end{eqnarray}
where the $Z$'s are matching (renormalization) factors which explicit
expressions up to one--loop order are obtained by combining the results of steps 1
and 2 in the Introduction (we refer the reader to ref.\cite{gm}
and references therein, 
where all the relevant formula are collected \footnote{In eq.(19) of
ref.\cite{gm} there is a misprint: the correct value of 
$\hat{\gamma}^{(0)}_{22}$ is $\hat{\gamma}^{(0)}_{22}=-8/3$.}) and 
eq.(\ref{eqocon}) which corresponds to step 3 in the Introduction. 
The values of the constants $D$'s which enter eq.(\ref{eqocon})
are given in Table \ref{numdo}.
Notice that in all previous Clover HQET computations of $B_{B}$, 
the incorrect value of the constant $D_{R}=D_{RR}+D_{RR}^{I}+D_{RR}^{II}=-5.4$
was used, to be compared to the correct one $D_{R}=-3.23$.
The physical value of the B--parameter is then given by
\be
B_{B}(\mu)\, =\, \sum_{i=LL,RR,N,S}\, Z_{O_{i}}(\mu)\, Z_{A}^{-2}(\mu)\, 
\mfrac{\langle \bar{B}^{0} \mid
O_{i}^{lat}(a)\, \mid B^{0} \rangle}{\mfrac{8}{3}\, \mid\langle 0 \mid
A_{4}^{lat}(a) \mid B^{0} \rangle \mid^{2}}\, \equiv\, \sum_{i}\, B_{O_{i}} 
\ee
The numerical determination of the relevant lattice matrix elements, step 4 in
the Introduction, is also discussed in detail in ref.\cite{gm}. We will take the 
values quoted in  ref.\cite{gm} as input for our reanalysis.

We now turn to the delicate point of the improvement of our lattice
perturbation theory results.
We start by discussing the choice of the expansion parameter.
Many different, but equivalent at one--loop level, definitions of the lattice 
coupling constant can be used in the perturbative calculation of the 
lattice--continuum matching.
At scales $a^{-1}\approx 2-4$ GeV of our numerical
simulations, we expect small non--perturbative effects in the renormalization
constants because of asymptotic freedom. However, tadpole diagrams, 
which appear in lattice perturbation
theory, give rise to large corrections and then to large uncertainties
in the matching procedure. In refs.\cite{parisi,boosted}, some recipes, usually refer 
to as Boosted Perturbation Theory (BPT), for choosing an optimal lattice coupling
constant which would absorb many of the unwanted gluonic tadpole contributions
are suggested. The claim is that the perturbative estimate of the
renormalization constants obtained from BPT at low orders, is closer to the
non--perturbative result than its Standard Perturbation Theory (SPT)
counterpart. Unfortunately, there are several different prescriptions 
which application is, in some cases, ambiguous.
In this work, we will calculate the renormalization constants for the
B--parameter with the following definitions of the coupling constant:

\begin{description}
\item[\bf SPT:] Standard Perturbation Theory, $\alpha^{SPT}_{s}=g^{2}/(4\, \pi)=
6/(4\, \pi\, \beta)$. It is well known that this coupling leads to poor
convergent perturbative series due to tadpole effects \cite{parisi,boosted}. We will use
this coupling only for the sake of comparison.
\item[\bf NBPT($\Box$):] Na\"{\i}ve Boosted Perturbation Theory based on the 
plaquette; $\alpha^{NBPT(\Box)}_{s}= 
\alpha^{SPT}_{s}/u_{0}^{4}$, where $u_{0}=(1/3\,  Tr(U_{\Box}))^{1/4}$, was
first introduced by Parisi \cite{parisi}. In our simulation at $\beta=6.0$, 
$1/3\, Tr(U_{\Box})=0.5937$.
\item[\bf NBPT($\kappa_{c}$):] Na\"{\i}ve Boosted Perturbation Theory based on the 
the critical Wilson hopping parameter, $\kappa_{c}$; $\alpha^{NBPT(\kappa_{c})}_{s}=
\alpha^{SPT}_{s}/u_{0}^{4}$, where $u_{0}=(8\, \kappa_{c})^{-1}$. In our
simulation at $\beta=6.0$, $\kappa_{c}=0.14543$ \cite{gm}.
%\item[\bf BPTD($\Box$):] Boosted Perturbation Theory with Tadpole Improvement
%based on the plaquette; $\alpha^{BPTD(\Box)}_{s}=
%\alpha^{NBPT(\Box)}_{s}$ and we tadpole improve the lattice operators
%\cite{boosted} as described in ref.\cite{wittig}.
%\item[\bf BPTD(${\cal K}_{c}$):] Boosted Perturbation Theory with Tadpole Improvement
%based on $\kappa_{c}$; $\alpha^{BPTD(\kappa_{c})}_{s}=
%\alpha^{NBPT(\Kappa_{c})}_{s}$ and we tadpole improve the lattice operators
%\cite{boosted} as described in ref.\cite{wittig}.
\item[\bf BPT($\ln(\Box)$):] Non--perturbative Boosted Perturbation Theory 
based on the plaquette's logarithm, $\ln(\Box)$. Following ref.\cite{boosted}, 
to obtain the coupling $\alpha^{BPT(\ln(\Box))}_{s}(q^{*})$ 
\begin{enumerate}
\item we solve for $\alpha_{V}(3.41/a)$ the expansion of the plaquette's 
logarithm
\[
\ln ( \mfrac{1}{3}\, Tr(U_{\Box}) )\, =\, -\mfrac{4\pi}{3}\, \alpha_{V}(3.41/a)\,
\left[\, 1\, -\, (1.19\, +\, 0.025\, n_{f})\, \alpha_{V}(3.41/a)\, \right];
\]
\item then we evolve $\alpha_{V}(3.41/a)$ down to the Lepage--Mackenzie's 
optimal scale $q^{*}$ (see below) through the two--loop renormalization group equation with the number 
of flavours $n_{f}=0$ because our simulation is performed in the quenched
approximation.
%\item The method to find the scale $q^{*}$ for two--fermion operators
%is described in ref.\cite{boosted}. 
\end{enumerate}

\end{description} 

Having defined the coupling, it remains to fix the scale at which it will be 
evaluated. Since $\alpha^{SPT}_{s}$, $\alpha^{NBPT(\Box)}_{s}$ and
$\alpha^{NBPT(\kappa_{c})}_{s}$ do not run, no scale setting is necessary in
contrast to the Lepage--Mackenzie's $\alpha^{BPT(\ln(\Box))}_{s}(q^{*})$.
Taking into account that $q^{*}$ is only meant to be a typical scale of
the process under consideration, we may guess the scale through some physical
arguments or, when the two--loop contributions are known, choose $q^{*}$ so that
the one--loop coefficient of the perturbative expansion vanishes. However, 
Lepage and Mackenzie have suggested a prescription to set $q^{*}$ which
consists in calculating the expectation value of $\ln(q^{2})$ in the one--loop
perturbative lattice contribution (the integrand of the integral 
defining the coefficient in front of $g^{2}/16
\pi^{2}$ in eqs.(\ref{eqocon}) to (\ref{eqocon4}) and eqs.(\ref{eqqcon})
to (\ref{eqqcon4}) ). It is clear that for divergent ($\mu$ dependent)
renormalization constants, $q^{*}$ depends on $\mu$. 
Hernandez and Hill \cite{hill}, however, have estimated $q^{*}$ for
heavy--light two--fermion operators using the Lepage--Mackenzie's prescription
without including the divergent terms proportional to $\ln(\mu^{2} \, a^{2})$.
Thus $q^{*}$ is independent of the renormalization scale $\mu$.
We will use this scheme to determine the optimal scale $q^{*}$ for
$Z_{A}$. Further, the  Lepage--Mackenzie's prescription is non ambiguous only 
if the operator under consideration does not mix with others under 
renormalization. In fact, if there is mixing, one can either calculate all 
renormalization constants at the same scale $q^{*}$ 
or determine a separate scale for each mixing coefficient \cite{vittorio}.
Since, in our case, the operators are divergent and do mix under
renormalization, we do not know how to implement the Lepage--Mackenzie's 
prescription. In view of these difficulties, 
we choose to determine $q^{*}$ for the heavy--light axial current
renormalization constant, $Z_{A}$, following ref.\cite{hill}, but for
four--fermion operators we will vary $q^{*}$ in the range 
$1\le a q^{*}\le \pi$, studying the dependence of $B_{B}$ on this scale.
In this study, for simplicity, we will use the same scale $q^{*}$ for all 
mixing coefficients. Moreover, we will use the results from the range of $q^{*}$ to
estimate a systematic error.

Following Lepage and Mackenzie, we now construct tadpole improved lattice operators 
by non--perturbatively redefining the hopping parameter $\kappa$, 
$\tilde{\kappa} = \kappa\, u_{0}$, where the mean--field parameter
$u_{0}$ can be taken to be the 
plaquette or $1/8 \kappa_{c}$. This redefinition leads to the rescaling 
of the standard normalization of lattice 
quark fields, $\sqrt{ 2 \kappa}$, with $\sqrt{u_{0}}$, $\sqrt{2 \kappa}\,
\sqrt{u_{0}}$, which is expected  
to remove tadpole contributions in the fermionic sector \cite{boosted}. 
Notice that tadpole contributions appear even when a good expansion parameter 
is used. To be consistent, the one--loop perturbative tadpole contributions must 
be subtracted from our perturbative 
expressions by multiplying by $1/\sqrt{u_{0}}$ expanded up to one--loop.
Moreover, all links operators $U(x)$ that appear in our lattice composite operators,
should be rescaled with $u_{0}$, $U(x)/u_{0}$, for the same reason as above.
This method, usually referred to as Tadpole Improvement (TD),
is expected to reorganize the perturbation series in such a way that
TD lattice operators have smaller discretization errors and the
lattice--to--continuum renormalization factors are nearer to unity. 
For the Wilson action,
TD has been studied in ref.\cite{boosted}. Its HQET counterpart is described
in refs.\cite{bernard,hill} to which we refer the reader for details. 
An important point to note is that for HQET quarks TD leads to a final result 
which is equivalent
to the reduction of the heavy--quark self--energy constant
$e$, already performed in the previous section. Therefore, heavy quark fields 
are already tadpole improved in our case.
Light quark fields, however, are not. As explained before, TD involves
multiplying all light--quark fields by 
$\sqrt{u_{0}}/\sqrt{u_{0}^{(1)}}$  
where $u_{0}^{(1)}$ is the one--loop perturbative estimate of the
mean--field parameter $u_{0}$. 
For example, consider a generic heavy--light four--quark operator $\tilde{O}$.
Its expression in terms of lattice operators is
\be \label{tdo}
\tilde{O}\, =\, (u_{0})\,
\left[\, 1\, +\, \mfrac{g_{s}}{16\pi^{2}}\, \left(\, D_{O}\, -\, 
D_{u_{0}}\, \right)\, \right]\, O^{lat\, TD}\, +\, \sum_{i}\, 
Z_{O_{i}}\, O^{lat}_{i}
\ee
where, to one--loop order, $u^{(1)}_{0}=1\, +\, \mfrac{g_{s}}{16\pi^{2}}\, D_{u_{0}}$.
For two--quark heavy--light operators, like the Axial current $A_{\mu}$,
$(u_{0})$ in eq.(\ref{tdo}) should be replaced with 
$\sqrt{u_{0}}$ and $D_{u_{0}}$ by $D_{u_{0}}/2$,
because now $\tilde{O}$ contains only one light field. We will implement TD
for both two and four--quark heavy--light operators and for all
choices of the coupling constant except for SPT.

As can be seen from eq.(\ref{tdo}), the value of $q^{*}$ for the tadpole
improved $Z_{A}$ depends on the choice of
the mean--field parameter $u_{0}$. We have used two definitions of $u_{0}$,
namely the plaquette and $1/8 \kappa_{c}$. Following ref.\cite{hill}, where the
calculation for the Wilson action was performed, and 
taking into account the contribution from the Clover action, we readily
obtain for $r=1.0$, $q^{*}_{A_{\mu}}(\Box)\, a=2.29$ for $u_{0}^{4}=1/3
Tr(U_{\Box})$ and $q^{*}_{A_{\mu}}(\kappa_{c})\, a=2.63$ for $u_{0}=1/8
\kappa_{c}$, to be compared to the Wilson
result $q^{*}_{A_{\mu}}(\kappa_{c})\, a=2.18$ \cite{hill}. 

Finally, we would like to discuss our estimation of the
systematic error due to the truncation of the perturbative series. 
The matching factors $Z$'s are the products of the renormalization factors
calculated in steps 1 to 3 in the Introduction, which have been computed
independently up to $O(\alpha_{s})$. In fact, the QCD operator $O_{LL}$ at the
scale $m_{b}$ can be written in terms of HQET operators as follows
\begin{eqnarray}
O_{LL}(m_{b}) &=& C_{L}(\mu)\, \tilde{O}_{LL}(\mu)\, +\,
C_{S}(\mu)\, \tilde{O}^{S}_{LL}(\mu)\nonumber\\
C_{L}(\mu) &=& 1\, +\, \mfrac{g^{2}}{16\, \pi^{2}}\, X_{L}(\mu)\nonumber\\
C_{S}(\mu) &=& \;\;\;\;\;\;\;\mfrac{g^{2}}{16\, \pi^{2}}\, X_{S}(\mu)
\end{eqnarray}
where $\mu$ is the renormalization scale in the HQET and the functions 
$X_{L}(\mu)$ and $X_{S}(\mu)$ have been
calculated at NLO in refs.\cite{twoloop} (see also \cite{gm,wittig}). The HQET
operators $\tilde{O}_{LL}$ and $\tilde{O}^{S}_{LL}$ can, in turn, be expressed
in terms of lattice HQET operators using eqs.(\ref{eqocon}) and (\ref{eqocon3}),
\begin{eqnarray}
\tilde{O}_{LL}(\mu) &=& 
\left(\, 1\, +\, \mfrac{g^{2}}{16\, \pi^{2}}\, Y_{LL}(a \mu)\right)\, 
O^{lat}_{LL}(a)\, +\, \mfrac{g^{2}}{16\, \pi^{2}}\, Y_{RR}\,
O^{lat}_{RR}(a)\nonumber\\
&+& \mfrac{g^{2}}{16\, \pi^{2}}\, Y_{N}\,
O^{lat}_{N}(a)\nonumber\\
\tilde{O}^{S}_{LL}(\mu) &=& 
\left(\, 1\, +\, \mfrac{g^{2}}{16\, \pi^{2}}\, Y^{S}_{LL}(a \mu)\right)\, 
O^{lat\: S}_{LL}(a)\, +\, \mfrac{g^{2}}{16\, \pi^{2}}\, \bar{Y}_{LL}(a \mu)\,
O^{lat}_{LL}(a)\nonumber\\
&+& \mfrac{g^{2}}{16\, \pi^{2}}\, \bar{Y}_{RR}\,
O^{lat}_{RR}(a)\, +\, \mfrac{g^{2}}{16\, \pi^{2}}\, \bar{Y}_{P}\,
O^{lat}_{P}(a)
\end{eqnarray}
Adding all together, we get the QCD--HQET lattice matching for $O_{LL}$,
\begin{eqnarray}\label{ollfull}
O_{LL}(m_{b}) &=& \left[\, 1\, +\, \mfrac{g^{2}}{16\, \pi^{2}}\,
(X_{L}\, +\, Y_{LL})\, +\, 
\left(\mfrac{g^{2}}{16\, \pi^{2}} \right)^{2}\, (X_{L} Y_{LL}\, +\, X_{S}
 \bar{Y}_{LL} )\, \right]\, O^{lat}_{LL}(a)\nonumber\\
&+& \left[\;\;\;\;\;\;\;\;\mfrac{g^{2}}{16\, \pi^{2}}\,
Y_{RR}\, +\, 
\left(\mfrac{g^{2}}{16\, \pi^{2}} \right)^{2}\, (X_{L} Y_{RR}\, +\, X_{S}
 \bar{Y}_{RR} )\, \right]\, O^{lat}_{RR}(a)\nonumber\\ 
&+& \left[\;\;\;\;\;\;\;\;\mfrac{g^{2}}{16\, \pi^{2}}\,
Y_{N}\, \; +\, \;
\left(\mfrac{g^{2}}{16\, \pi^{2}} \right)^{2}\, X_{L} Y_{N}\, \right]\, 
O^{lat}_{N}(a)\nonumber\\ 
&+& \left[\;\;\;\;\;\;\;\;\;\mfrac{g^{2}}{16\, \pi^{2}}\,
X_{S}\, \; +\, \;
\left(\mfrac{g^{2}}{16\, \pi^{2}} \right)^{2}\, X_{S} Y^{S}_{LL}\, \right]\, 
O^{lat\: S}_{LL}(a)\nonumber\\ 
&+& \left[\;\;\;\;\;\;\;\;\;\;\;\;\;\;\;\;\;\;\;\;\;\;\;\;\;\;\;\;\;\;\; 
\left(\mfrac{g^{2}}{16\, \pi^{2}} \right)^{2}\, X_{S} Y_{P}\, \right]\, 
O^{lat}_{P}(a)
\end{eqnarray}
with obvious notation. We can organize the renormalization constants in two
ways: including  $O(\alpha_{s}^{2})$ contributions
coming from the products of the factors expanded separately to $O(\alpha_{s})$
(method $M_{1}$) or excluding $O(\alpha_{s}^{2})$ terms in eq.(\ref{ollfull})
(method $M_{2}$).
Notice that the method $M_{1}$, in contrast to method $M_{2}$,
leads to results which depend on the renormalization scheme used in the 
definition of the intermediate continuum HQET operators. 
Since the lattice perturbative corrections are large,
different prescriptions may result in rather big discrepancies in the final value 
of $B_{B}(m_{b})$. We will view these differences as an estimation of the
systematic error on $B_{B}$ due to unknown higher order contributions to the
perturbative renormalization constants.
In refs.\cite{gm,wittig}, the method $M_{1}$ excluded terms of
$O(\alpha_{s}^{2})$ coming from the operator $\tilde{O}^{S}_{LL}$, 
i.e. they take $Y^{S}_{LL}=\bar{Y}_{LL}=\bar{Y}_{RR}=Y_{P}=0$ in 
eq.(\ref{ollfull}). This is inconsistent because there is no reason for 
excluding these contributions when
estimating the systematic error. Numerically the difference is very small
except for the contribution proportional to $O^{lat}_{P}$ due to the large
value of its matrix element between B--meson states.
This has been measured on the lattice using the same sample of 
gauge configurations as for $O_{LL}$, $O_{RR}$, $O_{N}$ and $O^{S}_{LL}$. We
refer the reader to ref.\cite{meulat97} for details. 

\begin{table}[t] \centering 
\begin{tabular}{||c|c|c|c|c|c|c|r|c||}
\hline
\hline
\multicolumn{3}{||c|}{Options}&\multicolumn{1}{c|}{$Z_{O_L}$}&
\multicolumn{1}{c|}{$Z_{O_R}$}&\multicolumn{1}{c|}{$Z_{O_N}$}&
\multicolumn{1}{c|}{$Z_{O_S}$}&\multicolumn{1}{c|}{$Z_{O_P}$}&
\multicolumn{1}{c||}{$Z_{A}$} \\
\hline
\hline
\multicolumn{2}{||c|}{$\alpha_s^{SPT}$} & $M_1$ &  0.7935& -0.0196& -0.0812& -0.1130&-0.0013&  0.9045\\ \cline{3-9}
\multicolumn{2}{||c|}{  0.0796} & $M_2$ &  0.7658& -0.0236& -0.1008& -0.1229& 0.0000& 0.8999\\ \hline
\multicolumn{2}{||c|}{$\alpha_s^{NBPT(\Box)}$} & $M_1$ &  0.7022& -0.0330& -0.1367&
-0.1063& -0.0023&  0.8145\\ \cline{3-9}
\multicolumn{2}{||c|}{  0.1341} & $M_2$&  0.6539& -0.0397& -0.1698& -0.1229& 0.0000& 
0.8067\\ \hline 
\multicolumn{2}{||c|}{$\alpha_s^{NBPT(\kappa_{c})}$} & $M_1$ &  0.6824& -0.0359&
-0.1487& -0.1048& -0.0025&  0.7950\\ \cline{3-9}
\multicolumn{2}{||c|}{  0.1458} & $M_2$ &  0.6297& -0.0432& -0.1847& -0.1229& 0.0000& 
0.7866\\ \hline
\multicolumn{2}{||c|}{$\alpha_s^{NBPT(\Box)}$ TD} & $M_1$&  0.7302& -0.0330& -0.1367&
-0.1084& -0.0023&  0.8312\\ \cline{3-9} 
\multicolumn{2}{||c|}{  0.1341} & $M_2$ &  0.7163& -0.0397& -0.1698& -0.1229&
0.0000&  0.8264\\ \hline
\multicolumn{2}{||c|}{$\alpha_s^{NBPT(\kappa_{c})}$ TD} & $M_1$ &  0.7446& -0.0359&
-0.1487& -0.1111& -0.0025&  0.8326\\ \cline{3-9}
\multicolumn{2}{||c|}{  0.1458} & $M_2$&  0.7387& -0.0432& -0.1847& -0.1229&
0.0000& 0.8282\\ \hline 
$q^{\ast}\! a$ & $\alpha_s^{BPT(\ln(\Box))}$ & \multicolumn{7}{|c||}{\mbox{}}\\ \hline
\mbox{} & \mbox{} & $M_1$ &  0.7092& -0.0603& -0.2497& -0.1148& -0.0041&  0.7993\\ \cline{3-9}
\raisebox{1.5ex}[0pt]{1.000} & \raisebox{1.5ex}[0pt]{  0.2449}& $M_2$ &  0.6977&
-0.0725& -0.3102& -0.1229& 0.0000&  0.7957\\ \hline
\mbox{} & \mbox{} & $M_1$ &  0.7061& -0.0396& -0.1636& -0.1111& -0.0030&  0.7993\\ \cline{3-9}
\raisebox{1.5ex}[0pt]{2.000} & \raisebox{1.5ex}[0pt]{  0.1812}& $M_2$ &  0.7099&
-0.0480& -0.2054& -0.1184& 0.0000&  0.7957\\ \hline
\mbox{} & \mbox{} & $M_1$ &  0.7026& -0.0370& -0.1528& -0.1104& -0.0028&  0.7993\\ \cline{3-9}
\raisebox{1.5ex}[0pt]{2.290} & \raisebox{1.5ex}[0pt]{  0.1727}& $M_2$ &  0.7083&
-0.0449& -0.1921& -0.1177& 0.0000&  0.7957\\ \hline
\mbox{} & \mbox{} & $M_1$ &  0.6986& -0.0347& -0.1430& -0.1098& -0.0027&  0.7993\\ \cline{3-9}
\raisebox{1.5ex}[0pt]{2.630} & \raisebox{1.5ex}[0pt]{  0.1648}& $M_2$ &  0.7060&
-0.0421& -0.1800& -0.1170& 0.0000&  0.7957\\ \hline
\mbox{} & \mbox{} & $M_1$ &  0.6944& -0.0327& -0.1347& -0.1093& -0.0025&  0.7993\\ \cline{3-9}
\raisebox{1.5ex}[0pt]{3.000} & \raisebox{1.5ex}[0pt]{  0.1580}& $M_2$ &  0.7033&
-0.0397& -0.1698& -0.1163& 0.0000&  0.7957\\ \hline
\mbox{} & \mbox{} & $M_1$ &  0.6929& -0.0320& -0.1320& -0.1091& -0.0025&  0.7993\\ \cline{3-9}
\raisebox{1.5ex}[0pt]{$\pi$} & \raisebox{1.5ex}[0pt]{  0.1557}& $M_2$ &  0.7023&
-0.0389& -0.1664& -0.1161& 0.0000&  0.7957\\ \hline
\hline
\end{tabular}
\caption{\it{Renormalization constants for different choices of the
lattice coupling constants, $q^{*}$ and options, see the text. The value of
$Z_{A}$ for the coupling $\alpha_s^{BPT(\ln(\Box))}$ has been calculated at
$q^{*}_{A_{\mu}}=2.63$. For this coupling, TD is computed using $u_{0}=1/8
\kappa_{c}$.}}
\label{tab:Ztable}
\end{table}

Now we give our final results.
In Tables 6 and 7, we present the numerical values of the renormalization constants
and the B--parameter $B_{B}(m_{b})$ for 
different choices of the coupling constant, the Lepage--Mackenzie's scale
$q^{*}$ and different ways in which we can
organize the perturbative series (methods $M_{1}$ and $M_{2}$) \cite{gm}. 
Our results have been obtained using $\alpha_{s}(\mu)$ at NLO, $m_{b}= 5$ GeV, 
$\Lambda^{n_{f}=4}_{QCD} = 200$ MeV and four active quark flavours, $n_{f}=4$.
As can be seen, $Z_{O_{P}}$ reduces the
value of $B_{B}$ from method $M_{2}$ by an amount smaller than the statistical
errors ($3$-$4$\%) and resulting in a reduction of the discrepancy between 
methods $M_{1}$ and $M_{2}$.
Further, the inclusion of the correct $D_{R}$ gives a value of $Z_{O_{R}}$
which is roughly one--half of the one used in refs.\cite{gm,wittig}.
Its non--negligible effect on $B_{B}$ is to increase its value by an amount 
slightly larger than the statistical errors (about $6$ -- $8$ \%). 
By comparing the results obtained without and with TD 
in options 2 to 5 of Tables 6 and 7 for NBPT couplings, it is clear that the 
Tadpole improvement of the operators significantly reduces the difference 
between methods $M_{1}$ and $M_{2}$ and hence the systematic error. In fact, 
when TD is implemented results from methods $M_{1}$ and $M_{2}$ are completely 
compatible within statistical errors. We have also implemented TD
for $\alpha^{BPT(\ln(\Box))}(q^{*})$ with two definitions of the mean--field 
parameter $u_{0}$: using the plaquette or the hopping parameter.
Again we found that the two definitions give results which are 
in perfect agreement within statistical errors.
Moreover, the values of $B_{B}$ obtained
with $\alpha^{BPT(\ln(\Box))}(q^{*})$ are almost independent of $q^{*}$ in the 
range $2\le q^{*} \le \pi$ and compatible within statistical errors with those
calculated with $\alpha^{NBPT(\Box)}$ and $\alpha^{NBPT(\kappa_{c})}$ for both 
methods $M_{1}$ and $M_{2}$. For this reason, we estimate our final value of 
the $B$--parameter for the coupling $\alpha^{BPT(\ln(\Box))}$ 
from the results in the interval $2\le q^{*} \le \pi$. Finally, we have made the
exercise of removing TD for this coupling. The main consequences are the 
following:
\begin{enumerate}
\item the value of the $B$ parameter significantly depends on $q^{*}$. 
For instance, from method $M_{1}$, $B_{B}$ varies from 0.74(5) at $q^{*} a=2$ to
0.86(5) at $q^{*} a=\pi$.
\item the discrepancy between methods $M_{1}$ and $M_{2}$ is much larger than 
with TD. For instance, from method $M_{2}$, $B_{B}=0.65(5)$ at $q^{*} a=2$
and $B_{B}=0.80(5)$ at $q^{*} a=\pi$, to be compared to the values from method 
$M_{1}$ given above.
\item the average value of $B_{B}$ is smaller than the TD one. 
\end{enumerate}

\begin{table}[t] \centering 
\begin{tabular}{||c|c|c|c|c|c|c|r|c||}
\hline
\hline
\multicolumn{3}{||c|}{Options}&\multicolumn{1}{c|}{$B_{O_L}$}&
\multicolumn{1}{c|}{$B_{O_R}$}&\multicolumn{1}{c|}{$B_{O_N}$}&
\multicolumn{1}{c|}{$B_{O_S}$}&
\multicolumn{1}{c|}{$B_{O_P}$}&\multicolumn{1}{c||}{$B_{B_{d}}$} \\

\hline
\hline
\multicolumn{2}{||c|}{$\alpha_s^{SPT}$} & $M_1$ & 0.912(43)& -0.023(1)& -0.097(6)& 
0.083(4)& -0.013(0)&  0.862(44)\\ \cline{3-9}
\multicolumn{2}{||c|}{ 0.0796} & $M_2$ & 0.889(42)&-0.027(1)&-0.122(7)& 0.091(4)&
0.000(0)& 0.831(43)\\ \hline
\multicolumn{2}{||c|}{$\alpha_s^{NBPT(\Box)}$} & $M_1$ &  0.995(47)& -0.047(2)&
-0.202(13)&  0.096(5)& -0.027(0)&  0.816(49)\\ \cline{3-9}
\multicolumn{2}{||c|}{ 0.1341} & $M_2$ & 0.945(45)&-0.057(2)&-0.256(16)& 0.113(6)&
0.000(0)& 0.745(48)\\ \hline
\multicolumn{2}{||c|}{$\alpha_s^{NBPT(\kappa_{c})}$} & $M_1$ & 1.015(48)& -0.053(2)&
-0.231(15)&  0.099(5)& -0.030(0)&  0.800(51)\\ \cline{3-9}
\multicolumn{2}{||c|}{ 0.1458} & $M_2$ & 0.957(45)&-0.066(3)&-0.293(19)& 0.119(6)&
0.000(0)& 0.718(50)\\ \hline
\multicolumn{2}{||c|}{$\alpha_s^{NBPT(\Box)}$ TD} & $M_1$ &  0.993(47)& -0.045(2)&
-0.194(12)&  0.094(5)& -0.026(0)&  0.823(49)\\ \cline{3-9}
\multicolumn{2}{||c|}{ 0.1341} & $M_2$ & 0.986(47)&-0.055(2)&-0.244(15)& 0.108(5)&
0.000(0)& 0.796(50)\\ \hline
\multicolumn{2}{||c|}{$\alpha_s^{NBPT(\kappa_{c})}$ TD} & $M_1$ & 1.010(48)& -0.049(2)&
-0.210(13)&  0.096(5)& -0.028(0)&  0.819(50)\\ \cline{3-9}
\multicolumn{2}{||c|}{ 0.1458} & $M_2$ &1.012(48)&-0.059(2)&-0.264(17)& 0.107(5)&
0.000(0)& 0.797(51)\\ \hline
$q^{\ast}\! a$ &$\alpha_s^{BPT(\ln(\Box))}$  & \multicolumn{7}{|c||}{\mbox{}}\\ \hline
\mbox{} & \mbox{} & $M_1$ & 1.043(49)& -0.089( 4)& -0.383(25)&  0.108( 5)& -0.050( 0)& 
0.629(56)\\ \cline{3-9}
\raisebox{1.5ex}[0pt]{1.0000} & \raisebox{1.5ex}[0pt]{ 0.2449}& $M_2$ & 1.036(49)&
-0.108( 5)& -0.480(31)&  0.116( 6)&  0.000( 0)&  0.564(59)\\ \hline
\mbox{} & \mbox{} & $M_1$ & 1.039(49)& -0.058( 2)& -0.251(16)&  0.104( 5)& -0.036( 0)& 
0.798(52)\\ \cline{3-9}
\raisebox{1.5ex}[0pt]{2.0000} & \raisebox{1.5ex}[0pt]{ 0.1812}& $M_2$ & 1.054(50)&
-0.071( 3)& -0.318(20)&  0.112( 5)&  0.000( 0)&  0.777(55)\\ \hline
\mbox{} & \mbox{} & $M_1$ & 1.034(49)& -0.054( 2)& -0.234(15)&  0.104( 5)& -0.034( 0)& 
0.814(52)\\ \cline{3-9}
\raisebox{1.5ex}[0pt]{2.2900} & \raisebox{1.5ex}[0pt]{ 0.1727}& $M_2$ & 1.052(50)&
-0.067( 3)& -0.297(19)&  0.112( 5)&  0.000( 0)&  0.799(54)\\ \hline
\mbox{} & \mbox{} & $M_1$ & 1.028(49)& -0.051( 2)& -0.219(14)&  0.103( 5)& -0.032( 0)& 
0.828(51)\\ \cline{3-9}
\raisebox{1.5ex}[0pt]{2.6300} & \raisebox{1.5ex}[0pt]{ 0.1648}& $M_2$ & 1.048(50)&
-0.062( 2)& -0.279(18)&  0.111( 5)&  0.000( 0)&  0.818(53)\\ \hline
\mbox{} & \mbox{} & $M_1$ & 1.022(48)& -0.048( 2)& -0.207(13)&  0.103( 5)& -0.031( 0)& 
0.839(51)\\ \cline{3-9}
\raisebox{1.5ex}[0pt]{3.0000} & \raisebox{1.5ex}[0pt]{ 0.1580}& $M_2$ & 1.044(50)&
-0.059( 2)& -0.263(17)&  0.110( 5)&  0.000( 0)&  0.833(53)\\ \hline
\mbox{} & \mbox{} & $M_1$ & 1.019(48)& -0.047( 2)& -0.202(13)& 0.102( 5)& -0.030( 0)& 
0.842(50)\\ \cline{3-9}
\raisebox{1.5ex}[0pt]{$\pi$} & \raisebox{1.5ex}[0pt]{ 0.1557}& $M_2$ & 1.043(49)&
-0.058( 2)& -0.258(16)&  0.110( 5)&  0.000( 0)&  0.837(53)\\ \hline
\hline
\end{tabular}
\caption{\it{Values of each operator contribution to the $B$--parameter, and
the total result, for different choices of the
lattice coupling constants, $q^{*}$ and options, see the text.}}
\label{tab:bbtable}
\end{table}

From Table 7, our best estimates of $B_{B}(m_{b})$ and 
the renormalization invariant $B$--parameter, $\hat{B}_{B}$, are 
\begin{eqnarray}\label{myfinalres}
B_{B}(m_{b}) &=& 0.81\, \pm \, 0.05\, \pm\, 0.03\, \pm\, 0.02\nonumber\\
\hat{B}_{B}(m_{b}) &=& 1.29\, \pm \, 0.08\, \pm\, 0.05\, \pm\, 0.03  
\end{eqnarray}
where the first error is statistical, the second is an estimate of the
uncertainty due to the choice of the coupling constant and the optimal scale
$q^{*}$ and the third takes into account the contribution of higher--order
terms to the renormalization constants.
We have computed $B_{B}(m_{b})$ using the NLO formulae.

We have repeated the same analysis with the lattice data for the $B$--$\bar{B}$
mixing from the UKQCD Collaboration obtained using the SW-Clover and HQET 
lattice actions at $\beta=6.2$ \cite{ukqcd}. The conclusions presented above
remain the same and the final results turn out to be in perfect agreement with
ours in eq.(\ref{myfinalres}),
\begin{eqnarray}
B_{B}(m_{b}) &=& 0.79\, \pm \, 0.04\, \pm\, 0.03\, \pm\, 0.02\nonumber\\
\hat{B}_{B}(m_{b}) &=& 1.26\, \pm \, 0.06\, \pm\, 0.05\, \pm\, 0.03  
\end{eqnarray}
Note also that we do not see any dependence of our results on the lattice
spacing $a$, within errors.

These numbers are to be compared with our previous result \cite{gm}:
$\hat{B}_{B}(m_{b})\, =\, 1.08\, \pm\, 0.06\, \pm\, 0.08$, obtained with the same
lattice data but with the incorrect values of the renormalization constants and
without tadpole improvement for the four--fermion operators.
Unlike our previous result, our best estimates in eq.(\ref{myfinalres}) are now in
good agreement with previous determinations of the $B$--parameter calculated 
by extrapolating Wilson data from the charm to the bottom mass 
(see Table 4 of ref.\cite{gm}).

\section{Conclusions}

In this paper we have determine the expressions of all $\Delta B=2$ and
$\Delta B=0$ four--fermion HQET continuum
operators in terms of HQET lattice ones using perturbation theory up to
one loop. This calculation is a necessary ingredient in the procedure to 
measure the unknown values of some important matrix elements of QCD 
four--fermion operators by simulating them with lattice HQET.
The calculation  has been performed in four continuum renormalization schemes,
NDR, DRED I, DRED II and HV, and for three lattice actions, Wilson, SW-Clover
with rotated fermion fields and SW-Clover with rotated propagators.
 
We have also discussed the subtleties associated with the renormalization
scheme dependence of our results. The effects of evanescent operators
which appear in intermediate steps of the computation have been identified and
clarified.

As a byproduct of our calculation, we have also obtained the one--loop
anomalous dimension matrices for both the $\Delta B=2$ and $\Delta B=0$  
HQET operators.

We have found and corrected an error in the one--loop result quoted in 
\cite{bp} for the lattice--continuum matching of the operator $O_{LL}$,
which matrix element between $B$--meson states determines the value
of the $B_{B}$ parameter of the $B$--$\bar{B}$ mixing.
Then we have reanalyzed the lattice data measured in ref.\cite{gm}
with the correct renormalization constant. We have used Boosted perturbation
theory with tadpole improvement to reduce the systematic uncertainties in the
renormalization constants due to the truncation of the perturbative series.
Our value for the $B_{B}$ parameter turns out to be 
larger and has a smaller systematic error than in
previous studies with static heavy quarks.  
Further, it is compatible with the results obtained by extrapolating lattice
Wilson data.

\section*{Acknowledgments}
We thank G. Martinelli for very useful discussions.
We also thank M. di Pierro and C. T. Sachrajda for helpful discussions.
We acknowledge the partial support by CICYT under grant number AEN-96/1718.
J.R. thanks the Departamento de Educacion 
of the Gobierno Vasco for a predoctoral fellowship.
 
\newpage
\myappendix
\appesection{Analytical expressions of the constants.}

In this appendix we give the analytical expressions for the lattice constants
which enter the continuum--lattice matching relations.   

\subsection*{$\Delta B=2$ operators.}

\noindent
{\bf Operator $O_{LL}$}

\begin{eqnarray}
D_{LL} &=& -1/3\, v\, -\, 10/3\, d_{1}\, -\, 1/3\, c\, -\, 4/3\, e\, -\,
4/3\, f\nonumber\\
D^{I}_{LL} &=& - 1/3\, v^{I}\, +\, 8/3\, (l+m)\, +\, 1/3\, s\, -\,
4/3\, f^{I}\, -\, 10/3\, n\nonumber\\
D^{II}_{LL} &=&  \delta_{L}\nonumber\\
D_{RR} &=&   4/3\, w\nonumber\\
D^{I}_{RR} &=&   4/3\, w^{I}\, -\, 2/3\, (l+m)\, +\, 1/3\, s\nonumber\\
D^{II}_{RR} &=&   \delta_{R}\nonumber\\
D_{N} &=& 2\, d_{2}\nonumber\\
D_{N}^{I} &=& -4\, d^{I}\, - \, 2\, q\, +\, 2\, h
\end{eqnarray}

\noindent
{\bf Operator $O_{LR}$}

\begin{eqnarray}
D_{LR} &=&   w\, +\, 1/6\, J_{1}\, -\, 7/3\, d_{1}\, +\, 1/6\, c
- 4/3\, e \, -\, 4/3\, f\nonumber\\
D^{I}_{LR} &=&  w^{I}\, +\, 13/6\, (l+m)\, +\, 1/4\, s\, +\, 1/6\, J_{2}
- 4/3\, f^{I}\, -\, 7/3 \, n\nonumber\\
D^{II}_{LR} &=&  3/4\, \delta_{R}\, +\, 1/6\, J_{3}\nonumber\\
\bar{D}^{S}_{RL} &=& 2/3\, w\, +\, J_{1}\, +\, 2\, d_{1}\, +\, c\nonumber\\
\bar{D}^{S\, I}_{RL} &=& 2/3\, w^{I}\, -\, 1/3\, (l+m)\, +\, 1/6\, s\, +\,
J_{2}\, +\, 2\, n\nonumber\\
\bar{D}^{S\, II}_{RL} &=& 1/2\, \delta_{R}\, +\, J_{3}\nonumber\\
D_{M} &=& d_{2}\nonumber\\
D^{I}_{M} &=& - 2\, d^{I}\, -\, q\, +\, h
\end{eqnarray}

\noindent
{\bf Operator $O^{S}_{LL}$}

\begin{eqnarray}
D^{S}_{LL} &=& -2/9\, v\, +\, 8/9\, J_{1}\, -\, 4/3\, d_{1}\, +\, 2/3\, c\, -\, 4/3\, e\, -\,
4/3\, f\nonumber\\
D^{S\, I}_{LL} &=& - 2/9\, v^{I}\, +\, 8/3\, (l+m)\, +\, 2/9\, s\, +\,
8/9\, J_{2}\, -\, 4/3\, f^{I}\, -\, 4/3\, n\nonumber\\
D^{S\, II}_{LL} &=&  2/3\, \delta_{L}\, +\, 8/9\, J_{3}\nonumber\\
\bar{D}_{LL} &=& 1/36\, v\, +\, 2/9\, J_{1}\, +\, 1/2\, d_{1}\, +\, 1/4\, c\nonumber\\
\bar{D}^{I}_{LL} &=&  1/36\, v^{I}\, -\, 1/36\, s\, +\, 2/9\, J_{2}
\, +\, 1/2\, n\nonumber\\
\bar{D}^{II}_{LL} &=&  - 1/12\, \delta_{L}\, +\, 2/9\, J_{3}\nonumber\\
\bar{D}_{RR} &=&  -  1/3\, w\nonumber\\
\bar{D}^{I}_{RR} &=&  -  1/3\, w^{I}\, +\, 1/6\, (l+m)\, -\, 1/12\, s\nonumber\\
\bar{D}^{II}_{RR} &=&  - 1/4  \delta_{R}\nonumber\\
D_{P} &=& - 1/4\, d_{2}\nonumber\\
D_{P}^{I} &=& 1/2\, d^{I}\, + \, 1/4\, q\, -\, 1/4\, h
\end{eqnarray}

\noindent
{\bf Operator $O^{S}_{LR}$}

\begin{eqnarray}
D^{S}_{LR} &=&   w\, +\, 1/6\, J_{1}\, -\, 7/3\, d_{1}\, +\, 1/6\, c
- 4/3\, e \, -\, 4/3\, f\nonumber\\
D^{S\, I}_{LR} &=&  w^{I}\, +\, 13/6\, (l+m)\, +\, 1/4\, s\, +\, 1/6\, J_{2}
- 4/3\, f^{I}\, -\, 7/3 \, n\nonumber\\
D^{S\, II}_{LR} &=&  3/4\, \delta_{R}\, +\, 1/6\, J_{3}\nonumber\\
\bar{D}_{RL} &=& 1/6\, w\, +\, 1/4\, J_{1}\, +\, 1/2\, d_{1}\, +\, 1/4\, c\nonumber\\
\bar{D}^{I}_{RL} &=& 1/6\, w^{I}\, -\, 1/12\, (l+m)\, +\, 1/24\, s\, +\, 1/4\,
J_{2}\, +\, 1/2\, n\nonumber\\
\bar{D}^{II}_{RL} &=& 1/8\, \delta_{R}\, +\, 1/4\, J_{3}\nonumber\\
D_{Q} &=& - 1/4\, d_{2}\nonumber\\
D^{I}_{Q} &=&  1/2\, d^{I}\, +\, 1/4\, q\, -\, 1/4\, h
\end{eqnarray}

\subsection*{$\Delta B=0$ operators.}

\noindent
{\bf Operator $Q_{LL}$}

\begin{eqnarray}
E_{LL} &=&  -\, 8/3\, d_{1}\, -\, 4/3\, e\, -\, 4/3\, f\nonumber\\
E^{I}_{LL} &=& 8/3\, (l+m)\, -\, 8/3\, n\, -\, 4/3\, f^{I}\nonumber\\
Et_{LL} &=&  J_{1}\, +\, 2\, d_{1}\, +\, c\nonumber\\
Et^{I}_{LL} &=& J_{2}\, +\, 2\, n\nonumber\\
Et^{II}_{LL} &=& J_{3}\nonumber\\
\bar{Et}^{S}_{RL} &=&   -\, 4\, w\nonumber\\
\bar{Et}^{S\, I}_{RL} &=&   -\, 4\, w^{I}\, +\, 2\, (l+m)\, -\, s\nonumber\\
\bar{Et}^{S\, II}_{RL} &=&   -\, 3\, \delta_{R}\nonumber\\
E_{R} &=& -\, 4/3\, d_{2}\nonumber\\
E_{R}^{I} &=& 4/3\, (2\, d^{I}\, +\, q\, -\, h)\nonumber\\
Et_{R} &=& -\, d_{2}\nonumber\\
Et_{R}^{I} &=& 2\, d^{I}\, +\, q\, -\, h
\end{eqnarray}

\noindent
{\bf Operator $Q_{LR}$}

\begin{eqnarray}
E_{LR} &=&  -\, 8/3\, d_{1}\, -\, 4/3\, e\, -\, 4/3\, f\nonumber\\
E^{I}_{LR} &=& 8/3\, (l+m)\, -\, 8/3\, n\, -\, 4/3\, f^{I}\nonumber\\
Et_{LR} &=&  v\, +\, 2\, d_{1}\, +\, c\nonumber\\
Et^{I}_{LR} &=& v^{I}\, -\, s\, +\, 2\, n\nonumber\\
Et^{II}_{LR} &=& -\, 3\, \delta_{L}\nonumber\\
Et_{RL} &=&   -\, 4\, w\nonumber\\
Et^{I}_{RL} &=&   -\, 4\, w^{I}\, +\, 2\, (l+m)\, -\, s\nonumber\\
Et^{II}_{RL} &=&   -\, 3\, \delta_{R}\nonumber\\
E_{S} &=& -\, 4/3\, d_{2}\nonumber\\
E_{S}^{I} &=& 4/3\, (2\, d^{I}\, +\, q\, -\, h)\nonumber\\
Et_{S} &=& -\, d_{2}\nonumber\\
Et_{S}^{I} &=& 2\, d^{I}\, +\, q\, -\, h
\end{eqnarray}

\noindent
{\bf Operator $Q^{S}_{LL}$}

\begin{eqnarray}
E^{S}_{LL} &=&  -\, 8/3\, d_{1}\, - \, 4/3\, e \, -\, 4/3\, f\nonumber\\
E^{S\, I}_{LL} &=&  8/3\, (l+m)\, - \, 4/3\, f^{I}\, -\, 8/3 \, n\nonumber\\
Et^{S}_{LL} &=&  4/3\, J_{1}\, -\, 1/3\, v\, +\, 2\, d_{1}\, +\, c\nonumber\\
Et^{S\, I}_{LL} &=&  -\, 1/3\, v_{I}\, +\, 1/3\, s\, +\, 4/3\, J_{2}\, +\, 2\, n\nonumber\\
Et^{S\, II}_{LL} &=&  \delta_{L}\, +\, 4/3\, J_{3}\nonumber\\
\bar{Et}_{LR} &=& -\, 1/3\, J_{1}\, +\, 1/3\, v\nonumber\\
\bar{Et}^{I}_{LR} &=& 1/3\, v^{I}\, -\, 1/3\, s\, -\, 1/3\, J_{2}\nonumber\\
\bar{Et}^{II}_{LR} &=& -\, \delta_{L}\, -\, 1/3\, J_{3}\nonumber\\
\bar{Et}_{RL} &=& -\, w\nonumber\\
\bar{Et}^{I}_{RL} &=& -\, w^{I}\, +\, 1/2\, (l+m)\, -\, 1/4\, s\nonumber\\
\bar{Et}^{II}_{RL} &=& -\, 3/4\, \delta_{R}\nonumber\\
E_{T} &=& -\, 4/3\, d_{2}\nonumber\\
E^{I}_{T} &=& 4/3\, (2\, d^{I}\, +\, q\, -\, h)\nonumber\\
Et_{T} &=& -\, d_{2}\nonumber\\
Et^{I}_{T} &=& 2\, d^{I}\, +\, q\, -\, h
\end{eqnarray}

\noindent
{\bf Operator $Q^{S}_{LR}$}

\begin{eqnarray}
E^{S}_{LR} &=&  -\, 8/3\, d_{1}\, -\, 4/3\, e\, -\, 4/3\, f\nonumber\\
E^{S\, I}_{LR} &=& 8/3\, (l+m)\, -\, 8/3\, n\, -\, 4/3\, f^{I}\nonumber\\
Et^{S}_{LR} &=&  J_{1}\, +\, 2\, d_{1}\, +\, c\nonumber\\
Et^{S\, I}_{LR} &=& J_{2}\, +\, 2\, n\nonumber\\
Et^{S\, II}_{LR} &=& J_{3}\nonumber\\
Et_{RR} &=&   -\, w\nonumber\\
Et^{I}_{RR} &=&   -\, w^{I}\, +\, 1/2\, (l+m)\, -\, 1/4\, s\nonumber\\
Et^{II}_{RR} &=&   -\, 3/4\,  \delta_{R}\nonumber\\
E_{U} &=& -\, 4/3\, d_{2}\nonumber\\
E_{U}^{I} &=& 4/3\, (2\, d^{I}\, +\, q\, -\, h)\nonumber\\
Et_{U} &=& -\, d_{2}\nonumber\\
Et_{U}^{I} &=& 2\, d^{I}\, +\, q\, -\, h
\end{eqnarray}
\begin{table}[!t] 
\begin{center}
\begin{tabular}{||c||r|r|r|r|r||}
\hline
\hline
\mbox{\rm Constant}&$r=1$&$r=0.75$&$r=0.50$&$r=0.25$&$r=0$\nonumber\\
%\mbox{\rm Constant}&\mbox{\rm Wilson}&\mbox{\rm Impr.~I}&\mbox{\rm Impr.~II}\nonumber\\
\hline
$J_{1}$&$-4.85$&$-4.93$&$-5.18$&$-6.08$&$-8.24$\nonumber\\
$J_{2}$&$-0.16$&$-0.21$&$-0.28$&$-0.25$&$0.00$\nonumber\\
$J_{3}$&$-0.11$&$-0.09$&$-0.06$&$-0.01$&$0.00$\nonumber\\
$\delta_{L}$&$0.20$&$0.15$&$0.09$&$0.02$&$0.00$\nonumber\\
$\delta_{R}$&$-1.23$&$-0.58$&$-0.17$&$-0.01$&$0.00$\nonumber\\
\hline
\hline
\end{tabular}
\caption{Numerical values of the new integrals for various values of
the Wilson parameter $r$.}
\label{int}
\end{center}
\end{table}
where the expressions and values for the constants $d_{1}$, $d_{2}$, $f$,
$v$, $w$, $c$ and $e$ which determine the matching in the Wilson case,
can be found in ref.\cite{flynn}. The Clover constants $l$, $m$, $n$, $q$,
$h$, $s$, $d^{I}$, $f^{I}$, $v^{I}$ and $w^{I}$ are defined in ref.\cite{bp}
whereas the analytical expressions of $\delta_{L}$ and $\delta_{R}$
are given in ref.\cite{frezzoti}.
We have found small numerical discrepancies between our estimates for 
$d^{I}$, $f^{I}$, $h$ and $v^{I}$ and the values quoted in ref.\cite{bp}.
Our results for $r=1.0$ are $d^{I}=-4.13$, $f^{I}=-3.64$, $h=-9.97$ and $v^{I}=-6.72$
to be compared to $d^{I}=-4.04$, $f^{I}=-3.63$, $h=-9.88$ and $v^{I}=-6.69$
from Table 3 of ref.\cite{bp}. We have used the former to obtain all our
results which agree with those recently and independently obtained in
ref.\cite{pierro}.

The only new constants are $J_{1}$, $J_{2}$ and $J_{3}$ defined as
\begin{eqnarray}
J_{1} &=& \mfrac{1}{\pi^{2}}\, \int_{\! -\pi}^{\, \pi} d^{4}k\left[\mfrac{\theta(1-k^{2})}{k^{4}} -
\mfrac{\Delta_{4}}{4\Delta_{1}\Delta_{2}^{2}} + \mfrac{\Delta_{5}}{4\Delta_{1}\Delta_{2}^{2}}
- \mfrac{r^{4} \Delta_{1}^{2}}{\Delta_{2}^{2}} - \mfrac{r^{2} \Delta_{4}}{2 \Delta_{2}^{2}}\right]\nonumber\\
J_{2} &=& \mfrac{1}{\pi^{2}}\,\int_{\! -\pi}^{\, \pi} d^{4}k\left[\mfrac{r^{2} \Delta_{4}^{2}}{8\Delta1\Delta_{2}^{2}} 
- \mfrac{r^{2}(\Delta_{4}-\Delta{5})}{2 \Delta_{2}^{2}}\right]\nonumber\\
J_{3} &=& \mfrac{1}{\pi^{2}}\,\int_{\! -\pi}^{\, \pi} d^{4}k\left[\mfrac{r^{4} \Delta_{4}^{2}}{16\Delta_{2}^{2}} 
- \mfrac{r^{4}\Delta_{1}(\Delta_{4}-\Delta{5})}{4 \Delta_{2}^{2}}\right]
\end{eqnarray}
where
\begin{eqnarray}
\Delta_{1} &=& \sum_{\mu}\, \sin^{2} \mfrac{q_{\mu}}{2}\nonumber\\
\Delta_{2} &=& \sum_{\mu}\, \sin^{2} q_{\mu}\, +\, 
4\, r^{2}\, \Delta_{1}^{2}\nonumber\\
\Delta_{4} &=& \sum_{\mu}\, \sin^{2} q_{\mu}\nonumber\\
\Delta_{5} &=& \sum_{\mu}\, \sin^{2} q_{\mu}\, \sin^{2} \mfrac{q_{\mu}}{2}
\end{eqnarray}
In Table \ref{int}, we give the numerical values of $J_{i}$, $i=1,2,3$
for several values of the Wilson parameter $r$. We also give the constants
$\delta_{L}$ and $\delta_{R}$ because in ref.\cite{frezzoti} they are not
presented explicitly.

\newpage

\begin{figure}

%\begin{center}
\setlength{\unitlength}{1mm}
\begin{picture}(20,180)
\put(0,0){\epsfbox{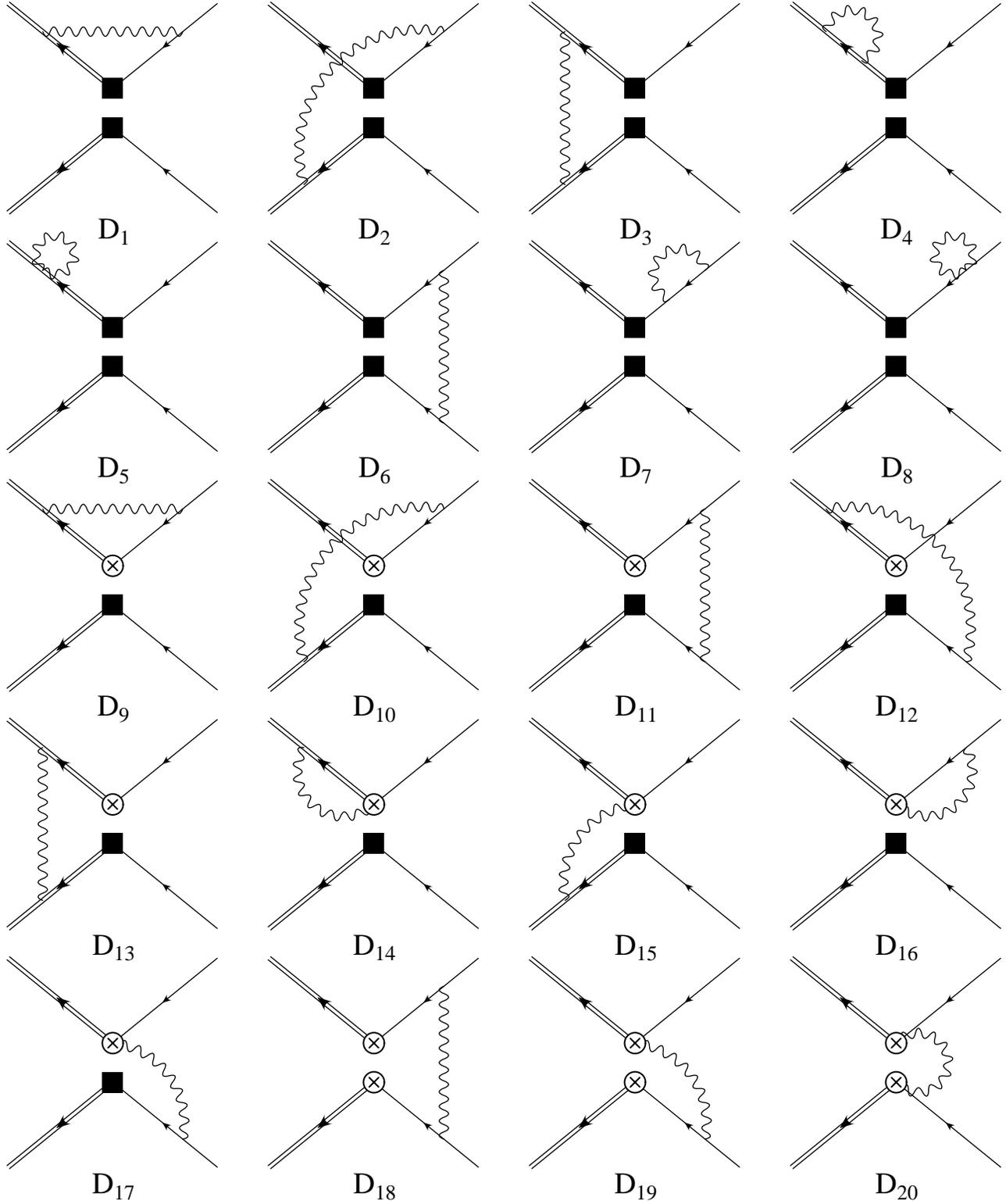}}
\end{picture}
%\end{center}
\caption{\it{Feynman Diagrams for the continuum--lattice matching. Up-down
replicas and $t$--channel diagrams are not shown. For the $\Delta B=0$ operators
only $s$--channel diagrams contribute, penguin diagrams are not considered and
the arrows in the lower fermionic lines must be inverted.}}
\label{fig:fourfig2}
\end{figure}

\end{document}